# Challenges and opportunities of ZnO-related single crystalline heterostructures

Y. Kozuka[1], A. Tsukazaki[2,3], M. Kawasaki[1,4,a)]

[1]*Department of Applied Physics and Quantum-Phase Electronics Center (QPEC), University of Tokyo, Tokyo 113-8656, Japan*

[2]*Institute for Materials Research, Tohoku University, Sendai 980-8577, Japan*

[3]*PRESTO, Japan Science and Technology Agency (JST), Tokyo 102-0075, Japan*

[4] *RIKEN Center for Emergent Matter Science (CEMS), Wako 351-0198, Japan*

## Abstract

Recent technological advancement in ZnO heterostructures has expanded the possibility of device functionalities to various kinds of applications. In order to extract novel device functionalities in the heterostructures, one needs to fabricate high quality films and interfaces with minimal impurities, defects, and disorder. With employing molecular-beam epitaxy (MBE) and single crystal ZnO substrates, the density of residual impurities and defects can be drastically reduced and the optical and electrical properties have been dramatically improved for the ZnO films and heterostructures with $Mg_xZn_{1-x}O$. Here, we overview such recent technological advancement from various aspects of application. Towards optoelectronic devices such as a light emitter and a photodetector in an ultraviolet region, the development of *p*-type ZnO and the fabrication of excellent Schottky contact, respectively, have been subjected to intensive studies for years. For the former, the fine tuning of the growth conditions to make $Mg_xZn_{1-x}O$ as intrinsic as possible has opened the possibilities of making *p*-type $Mg_xZn_{1-x}O$ through $NH_3$ doping method. For the latter, conducting and transparent polymer films spin-coated on $Mg_xZn_{1-x}O$ was shown to give almost ideal Schottky junctions. The wavelength-selective detection can be realized with varying the Mg content. From the viewpoint of electronic devices, two-dimensional electrons confined at the $Mg_xZn_{1-x}O$/ZnO interfaces are promising candidate for quantum devices because of high electron mobility and strong electron-electron correlation effect. These wonderful features and tremendous opportunities in ZnO-based heterostructures make this system unique and promising in oxide electronics and will lead to new quantum functionalities in optoelectronic devices and electronic applications with lower energy consumption and high performance.

[a)]Electronic mail: kawasaki@ap.t.u-tokyo.ac.jp

## TABLE OF CONTENTS

## I. INTRODUCTION

The aim of this review article is to overview the rapid progress of thin film techniques to grow ZnO based heterostructures on ZnO single crystalline substrates and to introduce novel physical properties and device functionalities that have become possible only for those heterostructures rather than conventional thin films of the same materials. There has been a long history of the research on Zinc Oxide (ZnO) as summarized in Fig. 1. Bulk form of ZnO such as ceramics and powders have been quite useful as chemical ingredients in rubbers, glasses, catalysts, pharmaceutical products like sunscreen, as well as magnetic ferrites and varistors as surge arrestor.[1] Basic properties of ZnO are summarized as follows: 1) a wide direct band gap of 3.37 eV, 2) a large exciton biding energy of 60 meV, 3) large spontaneous polarization and piezoelectric constants, 4) high purity, and 5) popular price and harmless as some of them are listed in Table I. For the electronics applications as passive components, most intensively studied topics are transparent conducting oxide films to be used for photovoltaics and displays.[2,3]

In the last decade or two, thin films studies have been rather focused on active device elements such as ultraviolet (UV) emitters and photodetectors as well as transparent transistors, with utilizing semiconducting features and transparency. Various kinds of reviews have been already reported, which highlight, for example, varistor application,[1] surface chemistry and physics,[4] defects,[5,6] ferromagnetism,[6,7] nanostructures,[6,8,9] Schottky junction,[10] UV emittion,[6,11] and transparent field-effect transistor.[2,3] One of the stimulating results triggering intensive studies aiming at such active device elements is the observation of highly efficient UV laser emission from thin films by photo excitation.[12,13] This experiments clearly indicated that ZnO thin films can be reached to the level of active semiconductor grade. For further exploring the possibility, it was evident that pursuing better thin film technique is essential to fabricate heterostructures as designed for active devices. Although there have been numbers of studies in this stream, we concentrate in this review article on the thin film techniques and devices on single crystalline ZnO substrates which are highlighted in blue characters in Fig. 1. The use of single crystalline substrates and molecular beam epitaxy (MBE) has drastically improved the performance of such devices as UV emitter, UV photodetector, and transistors. Especially two-dimensional (2D) electron transport with extremely high mobility has opened up a new research field on the quantum transport phenomena and possible quantum devices in oxides.

Many of the interesting functionalities originate from the Wurtzite crystal structure of ZnO as shown in Fig. 2. In this crystal structure, Zn is shifted from the center of octahedron with oxygen at vertices and the crystal does not possess inversion symmetry, showing a large spontaneous polarization along [0001] crystalline direction. Most studies on thin films have been focused on the Zn-polar (0001) or O-polar ($000\bar{1}$) due to energetic stability of the surfaces compared to other planes. Actually, thin films presented in this review article were mostly grown on Zn-plane ZnO substrate by molecular-beam epitaxy (MBE) due to the chemical stability against acid, high nitrogen doping efficiency, and proper direction of electrical polarization for creating 2D electron system. In Sections II and III, we introduce the growth optimization of ZnO and $Mg_xZn_{1-x}O$ thin films, respectively, on well-treated Zn-polar ZnO single crystal substrates. Then, we will discuss the various electrical or optical properties for the heteroepitaxial junctions. In Section IV, we describe photodetector comprised of conducting polymer/ZnO Schottky junction, which provides high quantum efficiency probably due to an abrupt and clean interface of organic-oxide constituents. In Section V, UV light emitter is overviewed based on *p*-type $Mg_xZn_{1-x}O$ / *n*-type ZnO junctions. In Section VI, 2D electron system confined at the $Mg_xZn_{1-x}O$/ZnO heterointerface is introduced as a potential candidate for the channel of quantum devices utilizing high electron mobility of 800,000 $cm^2\ V^{-1}\ s^{-1}$. Finally in Section VII, summary and future prospects are given for ZnO devices.

## II. HOMOEPITAXIAL GROWTH

### A. Reduction of impurities in ZnO substrate

Highly pure ZnO single crystals have been grown by vapor transport and hydrothermal methods.[14-16] In early stage of ZnO studies from 1960's to 90's, excitonic features had been intensively investigated by optical measurements of high quality single crystals.[17-21] In addition, surface charge transport had also been studied based on single crystal surfaces with electrolyte or ion implantations of He or H.[22-24] However, it had been difficult to apply ZnO single crystal substrates to thin film growth until around 2003 due to their high cost, short supply, Li contamination, and so on. In addition, it was necessary to develop a method making the surface of ZnO single crystal substrates epi-ready and atomically flat.[25]

Recently, hydrothermal method has been extended to produce large size (2 inch diameter) ZnO single crystals wafers (Fig. 1).[15,16] Prior to the chemical treatment described below, a high temperature annealing is applied to diminish Li contamination from ZnO wafers.[26] The surface of the single crystals can be mechano-chemically polished to realize apparently atomically flat surface. However, the as-polished and rinsed surface still has problems. We found that there remain abrasive powder $SiO_x$ particles on the surface, possibly due to the formation of $Zn(OH)_2$ gel.[25] We examined the electrochemical zeta-potential of abrasive powder and ZnO in the solutions with various pH. Setting the solution pH to give the same polarity for both materials, the reattachment of the particles to the ZnO surface can be avoided. Figure 3(a) shows the surface morphology of Zn-polar ZnO substrate etched by a HCl solution (HCl : $H_2O$ = 7 : 200) for 30 seconds, measured by an atomic force microscope (AFM). As shown by the magnified image in the inset, step and terrace structures are clearly formed. As an evidence for the elimination of Si contamination at the surface, secondary ion mass spectroscopy (SIMS) profile of Si was measured around the interface as shown in Fig. 3(b), exhibiting dramatic reduction compared with the case for as-received substrate. In addition, residual Li is also undetectable after high temperature annealing (not shown). Therefore, high temperature annealing and HCl etching are essential to fabricate high-quality films on single crystal substrates. It should be noted that the chemical resistance depends on the surface crystal planes, meaning Zn-polar, O-polar, or nonpolar M-surface presents different etching properties against acidic agents.

Second major impurity source was those coming from the substrate holder. We used to use a nickel-based alloy called Inconel® because it is known as an excellent alloy that is oxidation-resistive even at high temperatures.[27] In Fig. 4(a), the depth profile of SIMS for Mn is depicted for one of the samples grown using an Inconel® holder.[28] It is obvious that the Mn contamination exist in the film as well as around the interface. The origin for the increase of Mn concentration around interface is due to the surface contamination of Mn, probably during the annealing of the substrate in vacuum chamber prior to the growth. We then took quartz as the material for the substrate holders to eliminate any transition meal impurities, resulting in pure ZnO films free from such contamination elements as Ni, Mn, Co, and Mo under SIMS inspection.

Third possible impurity source originates from the oxygen radical source. We used to grow ZnO or $Mg_xZn_{1-x}O$ thin films by MBE equipped with oxygen radical source operated at 300W radio frequency, while the vapor fluxes of Zn (7N purity) and Mg (6N purity) were supplied from conventional Knudsen cells.[29,30] Even though the substrate etching process was established, we occasionally detected Si contamination, especially for the films grown at lower temperatures. In view of the high vapor pressure of SiO and the fact that radical source employs a crucible made of synthesized quartz, we suspected the crucible in oxygen radical source gun as a possible contamination source. Figures 4(b) and 4(c) represent the depth profiles of Si in the ZnO films grown at 850 °C and 920 °C, respectively, under otherwise the same conditions to yield in atomically flat surfaces. In the case of the higher growth temperature of 920 °C, re-evaporation of SiO may effectively reduce the Si contamination in the film [Fig. 4(c)]. Recently, we have changed the gas source from oxygen plasma to distilled pure ozone to supply contamination-free oxidizing agent.[31]

### B. Growth conditions of ZnO films

II/VI flux ratio is an important parameter for the growth of ZnO epitaxial films by MBE.[30-33] We varied the ratio of $Zn/O_2$ flux while keeping $T_g$ at 800 °C to tune the condition from O-rich through stoichiometric to Zn-rich conditions. Here, we used the ratios of $Zn/O_2$ = 3, 1, and 0.5, corresponding to O-rich, stoichiometric, and Zn-rich, respectively. The PL spectra taken at 12 K for those ZnO films are shown in Fig. 5 with the corresponding AFM images in the insets. The surface of all the samples remained atomically flat and there was no apparent influence of II/VI ratio, implying that $T_g$ is the major factor to obtain atomically flat surface. Photoluminescence (PL) spectra for all the samples clearly exhibited the free exciton absorption peak at 3.378 eV. However, peak at 3.420 eV assignable to the first excited state ($n$ = 2) of free exciton absorption[34] was observed only for O-rich [Fig. 5(a)] and stoichiometric [Fig. 5(b)] films. This result indicates that the higher optical quality was realized under stoichiometric or O-rich conditions. Note that the optimum II/VI ratio should be independently investigated at each $T_g$ due to the variation of vapor pressure of cations.

## III. GRWOTH OF MG$_x$ZN$_{1-x}$O FILMS

Given the high-quality single crystal substrate and well-developed growth technique of ZnO films, one of the remaining issues is the band gap engineering in order to fabricate, for example, double heterostructure light emitting diode. In general, the band gap of semiconductors can be varied by isovalent substitution through the change of lattice constants as Al doping give rise to an increase in the band gap of GaAs.[35] In the case of ZnO, band gap engineering can be achieved by substituting Mg or Cd with Zn, resulting in an increase or decrease in the band gap, respectively.[36-38] In this section, we overview the determination of Mg content in $Mg_xZn_{1-x}O$ films on ZnO substrate and the variation of physical properties as a function of $x$.[39]

### A. Determination of Mg content

There are several ways to determine the amount of specific elements. The appropriate method should be employed depending on the sensitivity and the precision required for the purpose. Another requirement is to discriminate the signals of the film from the substrate in the case of pseudomorphic growth on ZnO substrate. Under these considerations, we suggested using Rutherford backscattering spectroscopy (RBS) for high Mg concentration, and SIMS for low Mg concentration. Analytically quantified $x$ values in the $Mg_xZn_{1-x}O$ films are then used as the basis for calibration curves for the values of $c$-axis length deduced by X-ray diffraction (XRD) and localized exciton emission energy revealed by PL spectra. With using an almost linear relationship of these values with $x$, one can determine Mg content $x$ in $Mg_xZn_{1-x}O$ thin films on ZnO substrate with using standard laboratory techniques such as XRD and PL.

Figure 6(a) shows $\theta$–$2\theta$ diffraction patterns around the ZnO (0004) peak. For the high Mg region, a second peak, corresponding to the $Mg_xZn_{1-x}O$ layer is clearly observed together with Laue fringes which reflect the thickness of the film. The $c$-axis length difference ($\Delta c$) between the ZnO substrate ($c$ = 5.204 Å) and the $Mg_xZn_{1-x}O$ thin film is plotted in Fig. 6(b) for the four samples as a function of $x_{RBS}$ that was determined by RBS. The best-fit line was found to be $\Delta c$ (Å) = – 0.069 × $x$. This relation is significantly different from that for relaxed $Mg_xZn_{1-x}O$ films grown on $Al_2O_3$ substrates,[36] as a result of the films being under epitaxial strain in this case; the in-plane lattice is coherently connected with that of ZnO substrate and is contracted in comparison with the strain-free state. By using the relation shown in Fig. 6(b), interpolation gives estimates of Mg concentration, which we refer to as $x_{XRD}$, down to $x_{XRD}$ = 0.023. The lower bound is limited by the resolution of a conventional lab-based monochrometer in XRD equipment. Below this limit, the peak of the $Mg_xZn_{1-x}O$ film and ZnO substrate cannot be separated, as displayed in Fig. 6(a), for $x$ = 0.011. This limit may also be affected by the thickness or the quality of the film, which broaden the diffraction peaks.

To calibrate Mg concentration lower than $x \approx 0.02$, we focused on the exciton energy observed by PL as another physical parameter that almost linearly scales with Mg content. This method is conventionally used to determine Al composition in (Al,Ga)As thin films grown on GaAs substrate.[35] The energy dependence of the localized exciton luminescence on Mg concentration was investigated by PL at 100 K. The energy difference ($\Delta E$) between localized exciton emission from $Mg_xZn_{1-x}O$ layers and free exciton emission from ZnO is plotted in Fig. 6 as a function of $x$. The overall feature indicates that $\Delta E$ has a quite good linear dependence on Mg content particularly at low $x$ region as shown in Fig. 7(b). However, Fig. 7(a) indicates a nontrivial deviation of the localized exciton energies for high Mg content films toward lower energy from the extrapolated fitting line. This tendency is interpreted as stronger localization with higher Mg concentration. When the local Mg content fluctuates, the excitons tend to recombine after relaxing to lower band gap region. Thus, $c$-axis length from XRD is more appropriate to estimate Mg content at high Mg region. Unlike PL, XRD is relatively insensitive to the fluctuation in local Mg concentration due to its nature of reflecting the averaged structure. Therefore, we provide a fitted relation of $\Delta E$ (eV) = $2.2 \times x$, only valid for that of $x \leq 0.023$. This localized exciton energy dependence on Mg content is also similar to the case of $Mg_xZn_{1-x}O$ films grown on $Al_2O_3$ substrate.[36]

**B. Mg dependence of physical parameter of $Mg_xZn_{1-x}O$**

The precise $x$ value provided, we then examined physical properties dependent on Mg content. Figure 8(a) shows $x$ dependence of optical band gap ($E_g$) and $u$ parameter, where $u$ is estimated as $1/3(a/c)^2 + 1/4$ based on geometrical restriction (see Fig. 2) within the hard-sphere model. As $x$ increases, both the band gap and the $u$ parameter increases due to the deformation of crystal structure. Accordingly spontaneous polarization $P$ is also varied with Mg content, which is expressed as $\left[-8e/\left(\sqrt{3}a^2\right)\right]\left(u-3/8\right)$ with $q_1 \approx q_2 = -2e$. Spontaneous polarization of ZnO is about 5 $\mu C/cm^2$ according to first principles calculations,[40,41] and increases linearly with Mg content as shown in Fig. 8(b). As will be discussed later, this imbalance in $P$ will cause the discontinuity of polarization at the $Mg_xZn_{1-x}O$/ZnO interface, which is compensated by the charge accumulation at the interface. In Fig. 8(b), the estimated polarization discontinuity (solid line) and charge carrier density (solid circles) measured by Hall effect are compared, which indicates reasonable correspondence at low Mg region. At high Mg region, however, the measured value deviates from the linear relationship, which may suggest the presence of secondary contributions such as non-ideal deformation of crystal structure.

**C. Residual carrier density**

Residual charge carrier density ($N_D$) is one of important parameters that reflect the density of intrinsic defects in ZnO. We estimated $N_D$ in 1-μm-thick $Mg_xZn_{1-x}O$ ($0 \leq x \leq 0.39$) films grown on ZnO substrates with using Schottky junctions described in section IV.[42] The capacitance-voltage ($C$–$V$) relation was measured to deduce the depth profile of $N_D$ in the depletion layer of $Mg_xZn_{1-x}O$, as shown in the inset of Fig. 9 for an example. The arrow indicates the estimated $N_D$ value, while the steep increase to deeper side originates from the 2DEG accumulation at the heterointerfaces as described in Section VI. The $N_D$ data are plotted as a function of MgO molar fraction $x$ in Fig. 9 for the samples prepared under the optimized II/VI conditions at 920 °C. The $N_D$ in intrinsic $Mg_xZn_{1-x}O$ films stably lies in the range of $2 \sim 20 \times 10^{14}$ $cm^{-3}$ for $0 \leq x \leq 0.39$. Residual $N_D$ in undoped $Mg_xZn_{1-x}O$ films have been rarely reported in previous papers and accordingly the only thing already known about $N_D$ in conventional $Mg_xZn_{1-x}O$ films is in the order of $N_D \sim 10^{17}$ $cm^{-3}$ (Refs. 43,44). The present values of $N_D$ are not only the lowest ever reported but also even lower than those of undoped ZnO films with high electron mobility.[40] These results suggest that the $Mg_xZn_{1-x}O$ films grown on well-treated Zn-polar ZnO substrates by MBE are of electrically high quality in addition to the excellent optical properties[45] and hence the most important prerequisite has been satisfied for realizing *p*-type $Mg_xZn_{1-x}O$. For obtaining such a low $N_D$ in $Mg_xZn_{1-x}O$, we should investigate a dominant factor other than high $T_g$ as discussed above. A flux ratio is generally known to affect the MBE process of compound semiconductors, and thus the beam equivalent pressure (BEP) of Zn has to be adjusted so that Zn/O ratio is close to unity. We defined this stoichiometric BEP condition for given oxidation agent supply; the deposition rate linearly decreased below the stoichiometric BEP of Zn flux due to the shortage of Zn supplied onto a growing surface, while it stayed almost constant due to the evaporation of excess Zn from a surface above the stoichiometric BEP of Zn flux.[46]

## IV. SCHOTTKY CONTACT

Schottky junction is one the most fundamental unipolar semiconductor devices.[47] Oxide semiconductors are expectedly 'Schottky limit' due to large iconicity, where Schottky barrier height varies accordingly with the shift of work function of the metal.[48] However, it is not easy to obtain ideal Schottky behavior for ZnO with metals such as Au, Ni, and Pt without special surface treatment. Usually, ZnO junction with metals behaves as bad Ohmic-like contact that has rather large contact resistance without rectification by the Schottky barrier. In general, Schottky junction is applicable to the characterization of basic electrical properties of the depletion layers in the films or bulk crystals as shown in Fig.9 for example. Also, Schottky junctions are useful for a photodetector. To apply the benefit of Schottky junction on unipolar *n*-type ZnO, we need to develop a simple and reliable method for the fabrication of Schottky electrode. In this section, we describe the Schottky junctions composed of conducing polymer poly(3,4-ethylenedioxythiophene) poly(styrenesulfonate) (PEDOT:PSS) and $Mg_xZn_{1-x}O$ to demonstrate excellent UV light detectors.

### A. PEDOT:PSS/ZnO junction

We selected a conducting polymer, PEDOT:PSS as a candidate for Schottky electrode due to a large work function of 5.0 eV, a high conductivity at room temperature, and a simple process of spin coating. The ZnO surface was spin coated with PEDOT:PSS aqueous solution (H. C. Starck, Baytron PH500) and dried in a vacuum oven at 200 °C for 30 minutes in a glove box under Ar atmosphere. Figure 10(a) shows the chemical structures of PEDOT (left) and PSS (right). Polythiophene backbone of PEDOT forms the pathway for hole transport (π-conjugated system, dotted line), and PSS acts as an acceptor. The work function ($\phi_m$) of PEDOT:PSS was evaluated to be 5.0 eV from photoelectron yield spectrum,[49] which is similar to the value reported from other groups.[50-52] The film thickness of PEDOT:PSS was controlled by the rotation speed of spin-coating ranging from 40 nm to 80 nm with no significant change in the resistivity as low as ~ 1 mΩ cm. It is worth noting that the optical transmittance of a 50 nm thick PEDOT:PSS film showed excellent internal transmittance of > 90% for a wide region of wavelength from 250 nm to 1000 nm (not shown), which is suitable for optical application such as Schottky photodetectors. Figure 10(b) shows the schematic structure of the PEDOT:PSS/ZnO Schottky junction. Au contact electrode was deposited on the PEDOT:PSS/ZnO by thermal evaporation. Circular mesa structures with a diameter of 350 μm and ring-shaped Ohmic electrodes (Ti/Au) were formed on the ZnO substrate by Ar ion milling and electron-beam evaporation through photolithography technique, where the junction area was $9.6 \times 10^{-4}$ $cm^2$. The junction properties were examined by current density-voltage (*J*-*V*) and *C*-*V* measurements at room temperature in air under dark condition. The forward bias (positive *V*) corresponds to the current flow from PEDOT:PSS to ZnO. Figure 10(c) shows typical *J*-*V* characteristics in linear (right) and log (left) scales. The junction shows good rectifying behavior: the rectification ratio |*J* (2 V)|/| *J* (-2 V)| is as high as $\sim 10^7$. From the thermionic emission model,[47,53] a Schottky junction under forward bias has the *J*-*V* relation of

$$J = J_0 \exp\left(\frac{eV}{\kappa k_B T}\right) \qquad \left(V > 3k_B T / e\right), \tag{1}$$

where $J_0$ is the saturation current density, $e$ elementary charge, $V$ the applied voltage, $\kappa$ the ideality factor, $k_B$ Boltzmann's constant, and $T$ the absolute temperature. Here, $J_0$ is expressed as

$$J_0 = A^* T^2 \exp\left(-\frac{\phi_S}{k_B T}\right), \tag{2}$$

where $A^*$ is the effective Richardson constant with $A^* = 36$ A $cm^{-2}$ $K^{-2}$ for ZnO ($m^* = 0.29m_0$, $m_0$ the bare electron mass), and $\phi_S$ is the Schottky barrier height. The slope and the *J*-intercept from the linear fit to the semi-log plot for $V = 0.4 – 0.5$ V yield in $\kappa = 1.2$ and $\phi_S = 0.9$ eV, respectively. The $\kappa$ value close to unity indicates the high quality of the junction. The resultant rectification ratio and $\kappa$ are comparable to the best value using Pt and Ag electrode,[54-56] in spite of the simple spin-coating process in this study. The inset of Fig. 10(c) shows the histogram of $\phi_S$ variation among different 37 junctions on a ZnO substrate, indicating the considerably small deviation in their properties.

Now, we compare the Schottky junctions employing PEDOT:PSS and other noble metals in terms of $\phi_S$. The dependence of $\phi_S$ on work function ($\phi_M$) of the metal is characterized as $S$ parameter, which is expressed as[48,57]

$$S = \frac{\partial \phi_S}{\partial \phi_M} . \tag{3}$$

In the ideal case (Schottky limit), $S = 1$ is satisfied, which is mostly true of ionic semiconductors, while $S$ shows reduced values of ~ 0.1 in covalent semiconductors such as Si.[48] Figure 11 shows the relationship between $\phi_S$ and $\phi_M$-$E_C$ for the ZnO Schottky junction, where $E_C$ is the electron affinity.[10] Here, we distinguished between Zn-polar, O-polar, and other surfaces including M-surface and untreated surfaces. For low $\phi_M$-$E_C$ materials such as In, Ta, Al, and Ti, the junctions are Ohmic except Ag, indicating $\phi_S \approx 0$. The reason for the large $\phi_S$ for Ag may be due to interface oxide layer in the form of $AgO_x$.[58] As increasing $\phi_M$-$E_C$ to PEDOT:PSS, Au, and Pd, $\phi_S$ also increases, though there are scattered data points depending on the surface treatment. However, even large $\phi_M$-$E_C$ does not increase $\phi_S$ as in the case of Ir and Pt, which remains around $\phi_S$ = 0.8 eV. Based on Fig. 11, it is clear that the PEDOT:PSS/ZnO Schottky junction is almost on the line of $S = 1$, indicating clean and well-defined interface even though simple spin-coating process.

**B. Schottky photodetector**

For the application of UV photodetetor, transparency of the electrodes is required to introduce the UV light into the depletion layer in a semiconductor. Although the noble metals reflect the light from UV to infrared region, UV light can transmit through very thin metals. In contrast, of PEDOT:PSS has an excellent internal transmittance larger than 90% for a wide region of wavelength from 250 nm to 1000 nm.[59]

Figure 12 shows the spectral responses of the PEDOT:PSS/$Mg_xZn_{1-x}O$ Schottky photodiodes with various $x$.[60] As a fundamental value for evaluating the performance of a photodiode, the zero-bias responsivity, $R_\lambda = |I_{SC}|/A_{opt}P_\lambda = \eta e/hf$ (Ref. 20), is plotted as a function of wavelength of incident light, where $\eta$ is the quantum efficiency, $h$ is Planck's constant, and $f$ is the frequency of light. The line of $\eta = 1$ and $R_\lambda$ for the Si $p$-$i$-$n$ photodiode used in this study are also indicated for comparison. All the devices showed good UV/visible contrast of about $10^3$. The cut-off wavelength was shifted toward shorter wavelength with increasing $x$, reflecting the widening of the band gap energy ($E_g$) of $Mg_xZn_{1-x}O$. Above the $E_g$, all the curves exceeded that of the Si photodiode and mimicked the curve for ZnO independent of $x$, being in contrast with the situations for $Al_xGa_{1-x}N$ (Refs. 61, 62), where the maximum $R_\lambda$ decreases for higher Al contents. The maximum value of the $R_\lambda$ corresponds to the $\eta$ as high as about unity. In addition to the high transparency, an antireflection effect of 50-nm-thick PEDOT:PSS thin film seems one of the reasons for such high $\eta$ values observed in this study. Below the $E_g$, the $\eta$ decreased exponentially, following the Urbach's rule of $\eta \propto \exp(h\nu/E_{Urb})$, (Ref. 61) where $E_{Urb}$ is an experimental parameter named Urbach's energy. The steepness in the photo-response, i.e. $E_{Urb}$, was not largely degraded by alloying with Mg. It should be emphasized that the $x$ = 0.43 sample can selectively detect the light in UV-B region (280 nm - 315 nm) which is hazardous to human health.

## V. LIGHT EMITTER

Highly efficient light emitter at UV wavelength region opens a new door to various fields such as an alternative of fluorescent light or mercury lamp. GaN based light emitter has been recently widely prevails, not only blue but also white light-emitting diode has been expanded in use.[63,64] Towards UV light emitter based on ZnO, highly efficient emission is expected from the large exciton binding energy (60 meV). In addition, quantum well structure enhances the binding energy to 110 meV. This enhancement is significant not only because the excitons in quantum well are stable enough at room temperature but also because it surpass to the optical phonon energy (72 meV), being useful for laser operation.[65,66] Moreover, MBE method is one of the appropriate growth methods to mass production of commercial devices because of uniformity, purity, and large area.[67,68] Recent high-quality ZnO single crystal substrate with the conductivity of about a few S/cm enables us to fabricate vertical junction devices with low contact resistance.[69] To overcome the difficulty of synthesis for $p$-type ZnO, acceptor should be effectively doped to the films with low defect density in order to suppress the compensation effect. Therefore, we selected gas source $NH_3$ as an acceptor dopant during MBE growth at high temperature about 900 $^{o}$C.[67] Figure 13(a) shows a cross-sectional device structure of the vertical type light-emitting diode based on ZnO single crystal substrate. Semi-transparent 5nm thick Ni/Au top electrode was deposited on the $p$-type $Mg_xZn_{1-x}O$ layer for whole area and thick Ni/Au electrode for current correction on small area. Backside electrode to $n$-type ZnO substrate was prepared by In paste. Our electroluminescence measurement system was calibrated by using a commercial $In_xGa_{1-x}N$ based light-emitting diode (LED) having similar peak wavelength, whose output power was measured by a calibrated integrating sphere system. The output power ranged from 0.1 μW to 0.07 mW at the maximum attainable operation current (typically 30-40 mA). The LED was then coated with a 0.1 mm-thick epoxy resin containing 5 weight % $(BaEu)(MgMn)Al_{10}O_{17}$ (LP-G3, Mitsubishi Chemical) green phosphor. This structure is similar to the combination of commercial white LED composed of 470 nm blue (InGa)N LED and a garnet yellow phosphor, the only material choice with a high excitation efficiency by the blue LED. As shown in Fig. 13(b), a part of UV near-band-edge emission was converted into green emission. The UV emission of ZnO LED will make it possible to excite many existing phosphors developed for fluorescent tube, enabling better color rendering. Enhancing the hole concentration and reducing the contact resistance are the essential pathway for improving device performance.

## VI. TWO-DIMENSIONAL ELECTRON GAS

### A. Formation mechanism

2D electron gas (2DEG) has been conventionally created at semiconductor interfaces or surfaces utilizing various methods. Typical examples are electrostatic gating using a field-effect transistor structure as in $SiO_2$/Si (Ref. 70), or modulation doping as in AlGaAs/GaAs (Ref. 71). In contrast, the formation mechanism of 2DEG in $Mg_xZn_{1-x}O$/ZnO is dissimilar to these heterostructures in that no external filed or impurity doping is necessary; 2DEG is naturally present when the interface is formed. Figure 14 depicts how the 2DEG forms at the $Mg_xZn_{1-x}O$/ZnO interface. In the absence of charged adsorbents at the surfaces, spontaneous polarizations of ZnO and $Mg_xZn_{1-x}O$ generate an unphysically large internal electric field reaching ~ 5 MV/cm as shown in Fig. 14(a). This large electric field is usually compensated by surface adsorbents, although the exact potential relaxation mechanism is still controversial.[72-74] Irrespective of the surface charge compensation, uncompensated charges also remain at the $Mg_xZn_{1-x}O$/ZnO interface due to the difference in polarization between $Mg_xZn_{1-x}O$ and ZnO, which also leads to a large internal electric field. There are several ways to relax this unfavorable situation such as forming defects or intermixing of ions. If such disorder can be suppressed with improving the growth technique, the only way is to introduce additional charge carriers at the interface as shown in Fig. 14(b), which has been also verified in AlGaN/GaN interface.[75] Experimentally, we found that high-mobility electrons are accumulated at the interface as supported by the agreement between the carrier density estimated from Hall effect and that calculated from polarization difference as shown in Fig. 8(b).

### B. Depth profile of the 2DEG

Characterization of the depth profile of charge carriers in heterostructures is the first step to clarify the dimensionality of the accumulated charge carriers, which has conventionally been measured in semiconductors by *C-V* characteristics. Figure 15(a) shows a schematic device structure, where $Mg_xZn_{1-x}O/ZnO$ heterostructure was prepared on Zn-polar ZnO substrate by MBE. The Schottky junction was formed with PEDOT:PSS on the heterostructure as described in Section IV.[76,77] Figure 15(b) shows the depth profile of a representative device ($x$ = 0.05) in logarithmic (main) and linear (inset) scales. These plots are gathered with all data from iterative *V* scans in the same voltage range over 50 times. The intense peak of 2DEG is clearly revealed at $Mg_xZn_{1-x}O/ZnO$ heterointerface. The peak carrier concentration exceeds $1 \times 10^{19}$ cm$^{-3}$ and these carriers are accumulated at the ZnO layer adjacent to the $Mg_xZn_{1-x}O$ layer with the spreading width of about 5 nm, which is consistent to the value estimated from the triangular potential approximation.[78] By using the values of the peak carrier concentration of $1 \times 10^{19}$ cm$^{-3}$ and the spreading width of 5 nm, and assuming the triangular distribution of 2DEG, the sheet carrier concentration is roughly estimated to be ~ $10^{12}$ cm$^{-2}$. This value is consistent in order with the typical values of $Mg_xZn_{1-x}O/ZnO$ heterostructures evaluated by Hall coefficient and Shubnikov-de Haas oscillation periods.[79] The depth profile of charge carrier density evidences 2D conducting channel exists at the heterointerface, within 10-nm width. This measurements also indicate that the electrical bias from the surface can fully deplete the 2DEG accumulated at the interface, which means that the electron transport can be tuned on and off by the field-effect.

### C. 2DEG density tuning

Having established the formation of 2DEG with tunable density by Mg content $x$ as shown in Fig. 8(b), the transport characteristics of the 2DEG at the $Mg_xZn_{1-x}O/ZnO$ interface were measured with electrostatically varying the charge carrier density.[80,81] The first method we tried to modulate the carrier density is electrostatic gating. Here the Mg content is fixed at $x$ = 0.05. Figures 16(a) and 16(b) show a schematic diagram of the cross-sectional field-effect transistor structure and a plane view of the optical-microscope image, respectively. The gate electrode of Au/Ni was deposited on amorphous $Al_2O_3$. Output characteristics at 2 K for a field-effect transistor are presented in Fig. 16(c). The output curves show a clear saturation behavior due to pinch-off. The normally-on state with finite $I_D$ at $V_G$ = 0 V is due to the spontaneous accumulation of the 2DEG arising from the built-in electric field. The application of gate voltage ($V_G$) effectively modulated the drain current ($I_D$), exhibiting a large on/off ratio of approximately $10^5$ (not shown).[81] Leakage current ($I_G$) was suppressed below 100 pA in the entire range of $V_G$. The charge carrier density ($n$) in the 2DEG was evaluated as a function of $V_G$ from low-field Hall-effect measurements as shown in Fig. 16(d). All the traces appeared to be linear as a function of $V_G$ - $V_{th}$, where $V_{th}$ is the threshold voltages in the horizontal intercepts for $n$ - $V_G$ plot (not shown). Thus, the slope of this linear modulation behavior corresponds to the total capacitance of dielectrics in the device. In the repeated cooling and heating process, $V_{th}$ was slightly shifted due to charge trapping at the interfaces or within the dielectrics of $Al_2O_3$. While $V_{th}$ shifts slightly in a measurement at different temperatures, field-effect gating structure is reliable way to tune 2D transport with such polarization-induced 2DEG towards detailed investigations on quantum physics and transistor applications.

Another method for carrier density tuning is varying the Mg content. Although this method requires one heterostructure sample for each carrier density, there is an advantage in terms of electron mobility at low Mg content because Mg itself can be an additional source of scattering. We then focused on the transport properties in extremely low $x$ region below $x = 0.07$. The plotting of $\mu$ versus $n$ in Fig. 17 shows that the $\mu$ for diluted $x$ samples is apparently larger than the data (solid curve) taken for the field-effect transistor with relatively high Mg content ($x = 0.05$). From these data, we can infer the primary scattering mechanisms for either side of the highest mobility of $\mu = 770{,}000\ \mathrm{cm^2\ V^{-1}\ s^{-1}}$ at $n = 1.4 \times 10^{11}\ \mathrm{cm^{-2}}$ ($x \sim 0.01$) separately. At lower $n$, the points are aligned along $\mu \sim n$, which is a sign of the dominance of charged impurity scattering (CIS), suggesting more effective screening as the $n$ increases.[82] At higher $n$, i.e., as the Mg content is increased, the alignment of the points on the right hand side of the plot along $\mu \sim n^{-3/2}$ substantiates the interface roughness scattering (IRS) being the limiting mechanism.[83] Another consequence of the increased magnesium doping is a larger band offset at the interface, which affects the confinement potential of the 2DEG and the wave function is pushed towards the interface as $n$ increases, further enhancing IRS.[84]

**D. Electron mobility of the 2DEG**

Figure 18 summarizes the historical improvement of the transport scattering time $\tau_{tr}$ (left axis) and the mobility $\mu$ (right axis) of 2DEG for $Mg_xZn_{1-x}O$/ZnO heterostructures in comparison with that for AlGaAs/GaAs heterostructures.[85] Here, we plot $\tau_{tr}$ as a function of temperature as the difference in effective mass of electrons ($m^*$) is compensated by the relation of $\tau_{tr} = m^* \mu / e$. In the $Mg_xZn_{1-x}O$/ZnO system, the $\mu$ at low temperatures has been improved by two orders of magnitude in recent years through various kinds of improvements in the growth processes. The first quantum Hall effect (QHE) among oxides was demonstrated in 2007 for the 2DEG of $Mg_xZn_{1-x}O$/ZnO. The $\mu$ was as low as $5{,}500\ \mathrm{cm^2\ V^{-1}\ s^{-1}}$ because of the high density of unintentional impurities incorporated during the pulsed laser deposition (PLD) process.[69] Here, ZnO target could be very pure (impurity level of vapor-transport grown single crystals is well below one part per million[86]). However, $Mg_xZn_{1-x}O$ targets were prepared by solid state sintering in a furnace, resulting in the incorporation of ~ 0.1% impurities such as Si and Al (Ref. 87). Therefore, we switched the growth method from PLD to MBE in 2008. The heterostructures were grown on ZnO single crystal substrates from pure elemental sources of 7N Zn and 6N Mg. Oxygen radical ($O^*$) was supplied as an oxidizing agent through a radio frequency plasma source. By employing MBE, the $\mu$ was dramatically increased to 20,000 $cm^2\ V^{-1}\ s^{-1}$ (Refs. 69,88) and then to 180,000 $cm^2\ V^{-1}\ s^{-1}$ through careful optimization of the growth conditions.[28,77] Further improvement of the $\mu$ was realized by the installation of pure distilled ozone ($O_3$) source in the MBE system and by the careful optimization of the Mg content in the barrier layer;[31] the $\mu$ has reached to 770,000 $cm^2\ V^{-1}\ s^{-1}$ at $n = 1.4 \times 10^{11}\ cm^{-2}$ as discussed above. At present, the interface cleanness quantified by the value of $\tau_{tr}$ for ZnO system is comparable to that of AlGaAs/GaAs realized in 1986 exhibiting $\mu \sim$ 3,000,000 $cm^2\ V^{-1}\ s^{-1}$ (Refs. 89, 90). By the improvement of the $\mu$, various intriguing transport phenomena have been explored such as integer and fractional QHEs,[79,81] correlation effect,[88,91-94] and spin transport[95] as will be discussed later.

**E. Evolution of quantum Hall effect**

Figure 19 represents the magnetotransport data taken at each stage of the progress introduced above. The PLD sample shown in Fig. 19(a) was heteroepitaxially grown on the lattice matched $ScAlMgO_4$ substrates, resulting in O-polar growth.[79] Therefore, the sample has $ZnO/Mg_xZn_{1-x}O/ScAlMgO_4$-substrate structure in order to accumulate electrons at the $ZnO/Mg_xZn_{1-x}O$ interface. The other samples were grown on Zn-polar ZnO substrates by MBE resulting in $Mg_xZn_{1-x}O$/ZnO-film/ZnO-substrate structure. Although PLD grown samples had high concentration of impurities, the $\mu$ fulfills the criteria of Landau level quantization at a temperature of 0.045 K, exhibiting the Shubnikov-de Haas oscillations and quantized Hall plateaus. The sheet longitudinal resistance ($R_{xx}$) did not reach zero, probably due to the disorder in the 2D channel. By improving the $\mu$ to 140,000 $cm^2\ V^{-1}\ s^{-1}$ [Fig. 19(b)], the fractional QHE could be seen between $\nu = 1$ and $\nu = 2$. Not only at integer states but also at some fractional states such as $\nu = 5/3$ and 4/3, does $R_{xx}$ vanish at 0.06 K. When a high magnetic field was applied up to 27 T to a low $n$ sample, many fractional states such as $\nu = 2/3$, 2/5, and 1/3 were clearly observed beyond the quantum limit ($\nu = 1$ ) as shown in Fig. 19(c) (Ref. 87). More recently, the highest $\mu$ sample clearly shows a number of fractional Hall plateaus and dips in $R_{xx}$ at $\nu = 4/7$, 3/5, 5/11, and 4/9 (Ref. 31).

## F. Correlated 2DEG in $Mg_xZn_{1-x}O/ZnO$

In this section, we overview several phenomena that are characteristic of the 2DEG at the $Mg_xZn_{1-x}O/ZnO$ interfaces, where a large electron-electron correlation is expected due to its large effective mass. Thus far, a number of phenomena originating from the electron correlation have been suggested and experimentally verified in 2D systems as a representative example of fractional QHE. The strength of the electron correlation is quantified by Wigner-Seitz parameter $r_s$, which is the ratio of Coulomb energy [$e^2/4\pi\varepsilon(\pi n)^{-1/2}$] to kinetic energy ($\hbar^2 k_F^2/2m^*$, $k_F = \sqrt{2\pi n}$) and expressed as $\left(\sqrt{\pi n}a_B\right)^{-1} = (\pi n)^{-1/2} m^* e^2/4\pi\hbar^2\varepsilon$, where $\varepsilon$ is the dielectric constant and $a_B$ is the Bohr radius. In Fig. 20, a material-independent comparison of $r_s$ and transport scattering time $\tau_{tr}$ are plotted for $Mg_xZn_{1-x}O/ZnO$ heterostructures, the well-established GaAs,[89,96,97] Si/SiGe,[98] and GaN[83] 2DEGs. The larger $r_s$ in ZnO originates from the larger effective mass, which is distinct from the GaAs electron system. Note that the GaAs data are also shown for the most successful examples with larger $\tau_{tr}$, including the record-high $\mu$ of $36 \times 10^6$ cm$^2$ V$^{-1}$ s$^{-1}$ (Refs. 89,90). The largest $\tau_{tr}$ value of 120 ps at $r_s$ ~ 10 for ZnO corresponds to a mean free path of carriers as large as 3.2 μm. The most striking fact is that we can access the parameters region with $r_s > 10$ and moderately large $\tau_{tr}$, where new physics in 2D quantum transport may emerge in strongly correlated systems at clean oxide interfaces. Below, we focus on a representative aspects of electron correlation in ZnO 2DEG, which is an enhancement of spin susceptibility $g^*m^*$ ($g^*$ is the effective Landé $g$-factor) as a proximity to a ferromagnetic transition with renormalizing correlations based on Fermi liquid theory.

So far, spin susceptibility of a moderately correlated regime with $r_s > 1$ has been studied in a number of materials such as Si-MOSFET,[99-101] Si/SiGe,[102] GaAs/AlGaAs,[103,104] and AlAs/AlGaAs.[105,106] In these studies the enhancement of the effective spin susceptibility ($g^*m^*$) was universally observed. In order to evaluate $g^*m^*$ of ZnO 2DEG, we used the coincidence technique, where $R_{xx}$ was recorded under a magnetic field $B_{tot}$ at various tilt angles $\theta$ as the details explained in Ref. 84. The value of $m^*$ was also determined from the Dingle analysis, the temperature dependence of the Shubnikov–de Haas oscillations.[107] Figure 21(a) shows the carrier density dependence of the values of $g^*m^*$ and $m^*$ normalized to the bulk values of $g_bm_b$ and $m_b$, respectively, where $g_b = 1.93$ and $m_b = 0.29m_0$. Both $g^*m^*$ and $m^*$ increase with decreasing $n$, especially $g^*m^*$ shows a fourfold enhancement relative to $g_bm_b$. This result can be attributed to electron-electron interaction owing to the relatively large $r_s$.

Correlation effect is also renormalized into effective mass enhancement. To detect the renormalization, cyclotron resonance (CR) study has been carried out. This is because, for a translationally invariant isotropic electron gas, a long wavelength radiation can couple only to the center of mass of electron motion, which is not affected by the electron-electron interactions, giving the effective mass $m^*_{\mathrm{CR}}$ renormalized only by the electron-phonon interactions (Kohn's theorem[108]). On the other hand, the transport mass $m^*_{\mathrm{tr}}$ is additionally influenced by the electron-electron interactions due to the presence of the quasiparticle drag. Therefore, a comparison between $m^*_{\mathrm{CR}}$ and $m^*_{\mathrm{tr}}$ can be a unique tool of gauging the magnitude of the correlation effects. Figure 21(b) displays the $n$ dependence of the effective masses in the 2DEGs of $Mg_xZn_{1-x}O$/ZnO heterostructures, obtained by the present magnetotransport and CR measurements, $m^*_{\mathrm{tr}}$ and $m^*_{\mathrm{CR}}$, respectively. With reducing $n$, $m^*_{\mathrm{tr}}$ is steeply enhanced when $r_s > 5$ and exceeds $m_b$ by ~ 60% when $r_s \sim 10$. Similar mass enhancements at low $n$ have been observed in several 2DEGs such as AlGaAs/GaAs,[104] AlGaAs/AlAs,[109] and Si-based metal-oxide-semiconductor field-effect transistors.[110] In contrast, $m^*_{\mathrm{CR}}$ is nearly independent of the carrier density. Our results provide a strong evidence that the observed enhancement of $m^*_{\mathrm{tr}}$ compared to $m^*_{\mathrm{CR}}$ at the lowest carrier density originates purely from electron-electron interactions, owing to an $r_s$ as high as 10, which is difficult to achieve in other semiconductors. The value of $m^*_{\mathrm{tr}} / m^*_{\mathrm{CR}}$, which gauges the strength of the correlation effects, at the lowest $n$ in $Mg_xZn_{1-x}O$/ZnO is comparable to that of GaAs-based single-valley 2DEGs.[104,111]

It is challenging but intriguing to access the low carrier-density limit of $n < 1 \times 10^{11}$ cm$^{-2}$, which would further increase the correlation effects, while maintaining a large transport scattering time. One of the important consequences of electron correlation is spontaneous ferromagnetic transition, which is expected to be realized at $r_s > 25$ at zero field.[112] Although this transition has not been solidly verified in any 2D system, this property makes the ZnO heterostructures ideal systems for spintronic applications since polarized spins can be generated only by electron correlation with modulating carriers density.

**G. Spin coherence time**

Spintronics is one of the most attractive technologies towards low-energy devices or new quantum devices, which primarily focuses on the generation, manipulation, and detection of electron spins.[113-115] Transporting spin information is also a fundamental prerequisite. In this sense, ZnO 2DEG can be a promising candidate because spin coherence time is expected to be long because of extremely weak spin-orbit interaction[116] and of low concentration of nuclear spins as only 4 % of Zn isotopes possess a 5/2 nuclear spin.

In order to experimentally estimate these parameters, electrically detected electron spin resonance (ESR) has been carried out. In this technique, the resonance condition is detected though the change in resistance, leading to an estimation of Zeeman splitting energy, and hence *g*-factor of bare electrons without correlation effect. Spin relaxation time is also estimated from the full width at half maximum (FWHM) of the resonance peak. Upon the irradiation of microwave, sharp peaks were observed in $\Delta R_{xx}$ as shown in Fig. 22(a) at a resonant magnetic field ($B_{res}$) corresponding to ESR condition of $\hbar\omega = g^* \mu_B B_{res}$, where $\hbar$ is the Planck constant divided by $2\pi$, $\omega = 2\pi f$ is the angular frequency of the microwave ($f$ is the microwave frequency), and $\mu_B$ is the Bohr magneton.[117,118] In order to discuss the spin relaxation time, the ESR peaks are fitted by the Lorentzian function after subtracting the background as

$$\Delta R_{xx} = \frac{C_1}{(B - B_{res})^2 + C_2}, \tag{4}$$

where $C_1$ and $C_2$ are constants. Most of the ESR peaks are fitted well as shown in Fig. 22(a)

Collective transverse spin coherence time ($T_2^*$) was then estimated from the FWHM of the ESR peaks[118,119] as $T_2^* = 2\hbar / g^* \mu_B \Delta B$ ($\Delta B = 2\sqrt{C_2}$ is the FWHM of the resonance peak) from the fitting shown in Fig. 22(a). As other contributions such as longitudinal spin relaxation time ($T_1$) could affect the resonance line width, we take $T_2^*$ as the lower bound of the real transverse spin relaxation time ($T_2$) although $T_1 > T_2$ is always fulfilled and $T_2$ is usually the most dominant for the broadening. As shown in Fig. 22(b), $T_2^*$ is plotted as a function of $B_{res}$, exhibiting a maximum $T_2^*$ value of 27 ns and an enhancement at several magnetic fields. Such behavior is reminiscent of Shubnikov-de Haas oscillations as a function of $B_{res}$ (Refs. 120-122), however the enhancement does not exactly match the integer filling factors.

In Fig. 22(b), the expected spin relaxation time from the D'yakonov-Perel' process, which is due to Rashba spin orbit interaction, is also displayed as the solid curve $T_2$(Rashba), which is calculated by[123]

$$\frac{1}{T_2(\text{Rashba})} = \frac{\alpha^2 k_F^2 \tau_{tr}}{\hbar^2} \frac{2}{1+(\omega_L - \omega_c)^2 \tau_{tr}^2}, \quad (5)$$

where $\alpha$ is the Rashba spin orbit interaction parameter ($7.0 \times 10^{-14}$ eV m for this 2DEG) $k_F$ is the Fermi wavenumber, $\omega_L$ (= $\omega$ at the resonance) is the Larmor frequency, and $\omega_c$ is the cyclotron frequency. As shown in Fig. 22(b), $T_2$(Rashba) is roughly consistent with the experimental $T_2^*$ as an upper bound. However, there seems to be other limiting factors such as spin orbit interaction via momentum relaxation (Elliot-Yafet mechanism), electron-hole interaction (Bir-Aronov-Pikus), and the interaction with a small number of nuclear spins. Some of these mechanisms may also be comparatively dominant in our $Mg_xZn_{1-x}O$/ZnO heterostructure when the Rashba contribution is suppressed at high magnetic field.[121] Given the relatively long spin coherence time, ZnO 2DEG is an promising system for spintronic applications to transport and store the quantum information as spins, if continuing development of process technique to fabricate mesocsopic structures on ZnO heterostructures are made.

## VII. CONCLUSIONS

The quality of ZnO-related thin films and heterostructures has been drastically improved by the use of properly treated ZnO single crystal substrates and molecular-beam epitaxy. This review article has provided an overview of key technologies on the preparation of epi-ready substrate as well as the concept for optimization of growth conditions. For designing heterostructure devices and making a tight feedback to the growth, precise characterization of Mg content in the $Mg_xZn_{1-x}O$ films is extremely important but very few results have been reported for those grown on ZnO substrates. Here, we provide a calibration method to determine $x$ from the conventional measurements of PL and XRD, which will be useful for future research.

As minority carrier devices, we introduced optoelectronic devices such as UV emitters and detectors. Vertical configuration UV LEDs have been demonstrated with use of highly conducting ZnO substrates. The key challenge for making brighter LEDs are realizing higher hole density and forming better Ohmic contacts to *p*-type layer. In contrast, UV photodetector composed of PEDOT:PSS/$Mg_xZn_{1-x}O$/ZnO Schottky junctions have demonstrated a nearly ideal performance as wavelength selective sensors. Future challenge is to develop a diode laser in order to make full use of large exciton binding energy that will facilitate the laser action.

Majority carrier devices like field effect transistors on ZnO substrate have provided a unique opportunity for the research on the quantum physics of 2DEG such as quantum Hall effect. The rapid progress in realizing higher mobility starts to unveil unique physical properties originating from strong electron correlation effect. Due to its large spin susceptibility and very low Rashba spin orbit coupling, this system will pave ways for controlling and transpiring/storing spins, respectively. Further developments of quantum devices will lead to explore unrevealed quantum phenomena in mesoscopic structures such as quantum dots and quantum point contacts defined by electron beam lithography. Moreover, in comparison with conventional semiconductor 2D system, oxide heterostructures have tremendous degrees of freedom in material selection to form heterointerfaces. Therefore, spontaneous accumulation of 2D electrons at the heterointerface will explore new functionalities for electro-magnetic devices.

## ACKNOWLEDGMENTS

The authors gratefully acknowledge S. Akasaka, S. F. Chichibu, J. Falson, T. Fukumura, T. Makino, D. Maryenko, K. Nakahara, M. Nakano, A. Ohtomo, Y. Segawa, K. Tamura, K. Ueno, and H. Yuji. This work was partially supported by JSPS Grant-in-Aid for Scientific Research (S) No. 24226002 and for Young Scientists (A) (Grant No. 23686008), and by JST, CREST as well as by Asahi Glass Foundation. A. T. and M. K. were partly supported by the Japan Society for the Promotion of Science through the Funding Program for World-Leading Innovative R&D on Science and Technology (FIRST Program), initiated by the Council for Science and Technology Policy.

**Figure Captions:**

**FIG. 1.** (Color online) Historical stream from bottom to top for the research topics and applications of ZnO. Those in blue are discussed in this review manuscript. The inset pictures are (a) sintered ceramics, (b) powders, (c) a transparent transistor made on glass, (d) a blue light emitting diode made of ZnO *p-n* junction, (e) and atomic force microscope image of a ZnO single crystal substrate surface that exhibits step and terrace structure, and (f) a commercially available ZnO single crystal wafer with a diameter of 2 inches. [Adapted from Jpn. J. Appl. Phys. **44**, L643 (2005). Copyright 2005 The Japan Society of Applied Physics.]

**FIG. 2.** Wurtzite crystal structure of ZnO composed with solid and open circles for oxygen and Zinc atoms, respectively. Tetrahedron with four hold coordination plays a key role inside of the structure.

**FIG. 3.** (a) Atomic force microscope image of the etched surface of Zn-polar ZnO substrate. The magnified inset image exhibits straight step and terrace structure. (b) Depth profiles for Si measured with SIMS for the films grown on as-received and HCl-etched substrates. [Adapted from Appl. Phys. Express **4**, 035701 (2011). Copyright 2011 The Japan Society of Applied Physics.]

**FIG. 4**. The SIMS profiles of impurity for (a) Mn contamination in ZnO film grown at 980 $^{o}$C presumably originating from the Inconel® substrate holder, (b) Si contamination in ZnO film grown at 850 $^{o}$C presumably originating from oxygen plasma source. The quartz substrate holder was used for this sample. (c) Si signal is as low as the detection limit for the sample grown at 920 $^{o}$C. [Adapted from Jpn. J. Appl. Phys. **50**, 080215 (2011). Copyright 2011 The Japan Society of Applied Physics.]

**FIG. 5**. (Color online) Photoluminescence spectra measured at 12 K for the ZnO films grown at 800 $^{o}$C under (a) oxygen rich, (b) stoichiometric, and (c) Zinc rich conditions. The insets show surface morphologies measured by AFM corresponding to each sample. [Adapted from Jpn. J. Appl. Phys. **49**, 071104 (2010). Copyright 2010 The Japan Society of Applied Physics.]

**FIG. 6**. (Color online) (a) $\theta$-2$\theta$ X-ray diffraction around ZnO (0004) peak. The asterisks indicate the peaks corresponding to $Mg_{x}Zn_{1-x}O$ layers. The methods to determine $x$ are indicated for respective Mg concentration ranges as indicated. (b) *a*-axis length and *c*-axis length as a function of Mg content for $Mg_{x}Zn_{1-x}O$ films grown on $Al_2O_3$ (0001) substrates (triangles) and ZnO (0001) substrates (circles). [Reprinted with permission from J. Appl. Phys. **112**, 043515 (2012). Copyright 2012 AIP Publishing LLC.]

**FIG. 7.** (Color online) (a) The energy difference ($\Delta E$) between localized exciton emission from $Mg_{x}Zn_{1-x}O$ films and free exciton emission from ZnO as a function of $x$

at 100 K. The free exciton energy of ZnO is 3.368 eV at 100 K. The methods to determine $x$ are also indicated. The dashed lines are the fitting for the data below $x$ = 0.015 at 100 K. (b) Magnified view in the Mg content range of $x \leq 0.05$. [Reprinted with permission from J. Appl. Phys. **112**, 043515 (2012). Copyright 2012 AIP Publishing LLC.]

**FIG. 8.** (Color online) (a) Mg content dependence of the band gaps for $Mg_xZn_{1-x}O$ films grown on $Al_2O_3$ (0001) substrates and $u$ parameters for the films grown on ZnO (0001) substrates. (b) Calculated interface charges $\Delta P/e$ induced by the polarization mismatch between $Mg_xZn_{1-x}O$ layer and ZnO layer. Filled circles represent experimental data points for $Mg_xZn_{1-x}O$/ZnO heterostructures. The inset shows a magnified view in the low Mg range of $x \leq 0.05$.

**FIG. 9.** The relationship between ionized donor density ($N_D$) in undoped $Mg_xZn_{1-x}O$ and molar fraction $x$. The inset presents depth profile of $N_D$ in $Mg_xZn_{1-x}O$ grown on ZnO substrate. The bottom $N_D$ value indicated by arrow is plotted in the main panel. [Adapted from Appl. Phys. Express **3**, 071101 (2010). Copyright 2010 The Japan Society of Applied Physics.]

**FIG. 10.** (a) Chemical structures of PEDOT (left) and PSS (right). (b) The schematic cross section of the PEDOT:PSS Schottky junction fabricated on ZnO substrate. (c) The current density-voltage ($J$-$V$) characteristics for the PEDOT:PSS/ZnO Schottky junction. The histogram of $\phi_S$ among 37 junctions on a ZnO substrate is shown in the inset. [Reprinted with permission from Appl. Phys. Lett. **91**, 142113 (2007). Copyright 2007 AIP Publishing LLC.]

**FIG. 11.** (Color online) Schottky barrier height ($\phi_S$) as a function of $\phi_M - E_C$ for the Schottky junctions composed of ZnO and various metals, where $\phi_M$ is the work function of the metal and $E_C$ is the electron affinity of ZnO. Closed circles represent Zn-polar crystal, open circles O-polar crystals, and open triangles other surfaces such as M-surface or untreated one. Dashed line indicates the ideal Schottky barrier height.

**FIG. 12.** (Color online) Spectral response of the PEDOT:PSS/$Mg_xZn_{1-x}O$ Schottky photodetectors measured under a zero-bias condition at room temperature. The black and gray lines denote quantum efficiency equal to unity and the zero-bias responsivity for the Si *p-i-n* photodiode used in this study, respectively. The regions of UV-A, UV-V, and UV-C bands are also shown at upper axis. The inset shows quantum efficiency as a function of photon energy for the $x$ = 0.43 sample with the position of $E_{half}$. [Adapted from Appl. Phys. Express **1**, 121201 (2008). Copyright 2008 The Japan Society of Applied Physics.]

**FIG. 13.** (Color online) (a) Schematic diagram for the ZnO based light emitter grown

on ZnO substrate. Back electrode is indium and top electrode is thin Ni/Au electrode for transmitting the emission at the interface. (b) Electroluminescence spectra measured at room temperature with and without green phosphor. The inset shows a picture of light emission. [Reprinted with permission from Appl. Phys. Lett. **97**, 013501 (2010). Copyright 2010 AIP Publishing LLC.]

**FIG. 14.** (Color online) Schematic diagram of charge distribution and band diagram of the $Mg_xZn_{1-x}O$/ZnO heterostructure (a) before and (b) after surface and interface charges are compensated.

**FIG. 15.** (Color online) (a) A cross sectional schematic diagram of the device. (b) The depth profile of charge carrier density measured at 2 K. The inset shows the same plot in linear scale. [Reprinted with permission from Appl. Phys. Lett. **96**, 052116 (2010). Copyright 2010 AIP Publishing LLC.]

**FIG. 16.** (Color online) (a) Schematic cross-sectional view of the field-effect transistor with atomic-layer-deposited $Al_2O_3$ gate dielectrics. 2DEG is formed at the interface of $Mg_xZn_{1-x}O$/ZnO. (b) Optical microscope image of the device and measurement configuration. (c) Output characteristics at 2 K. (d) The relationship of mobility ($\mu$) and electron density ($n$) at 0.06 K (closed circles), 2 K (closed triangles), and 10 K (closed squares). The Mg content $x$ in this sample is ~ 0.05. The inset shows the $n$ as a function of the gate bias ($V_G$) in excess of the threshold voltage ($V_{th}$).

**FIG. 17.** Carrier density ($n$) versus electron mobility ($\mu$) relationship for the samples with various Mg content (circles) and by using electrostatic gating (solid curve). Broken lines on the left and right are guides to the eye for the limitations of charged impurity scattering and interface roughness scattering, respectively.

**FIG. 18.** (Color online) Comparison in historical improvement of transport scattering time $\tau_{tr}$ for ZnO and GaAs 2D electron gas. Closed and open circles represent the data for ZnO and GaAs, respectively. The $\tau_{tr}$ value is related to the electron mobility ($\mu$) shown in right axes as $\mu = e\tau_{tr} / m^*$, where $e$ is the elementary charge and effective mass of bulk electron. $m^*$ for ZnO and GaAs are $0.29m_0$ and $0.067m_0$, respectively.

**FIG. 19**. (Color online) Magnetoresistance $R_{xx}$ and Hall resistance $R_{xy}$ of $Mg_xZn_{1-x}O$/ZnO heterostructures with different mobility [$\mu$ ($cm^2$ $V^{-1}$ $s^{-1}$)] and carrier density [$n$ ($10^{11}$ $cm^{-2}$)]. The Landau indices are given at corresponding $R_{xy}$ plateaus. (a) The sample grown on $ScAlMgO_4$ substrate by pulsed laser deposition, which exhibited the first quantum Hall effect in oxides. The Mg content $x$ is 0.15. The samples were grown on ZnO substrates by MBE for (b), (c), and (d). (b) The sample with $x$ ~ 0.05 grown with using oxygen radical source. The samples with $x$ ~ 0.01 (c) and $x$ ~ 0.006 (d) were grown with using distilled pure ozone.

**FIG. 20**. (Color online) Mapping of electron–electron interaction parameter ($r_s$) and transport scattering time ($\tau_{tr}$) for 2DEGs of a variety of material systems. The data are taken from Refs 86,92,93 for GaAs electron systems ($0.067m_0$), from Ref. 94 for Si/SiGe ($0.22m_0$), and from Ref. 79 for GaN ($0.2m_0$).

**FIG. 21.** (Color online) (a) Spin susceptibility ($g^*m^*$) and effective mass ($m^*$) with respect to bulk values as a function of carrier density ($n$). Top axis shows the corresponding $r_s$. (b) The effective electron masses, which are obtained by the cyclotron resonance and Shubnikov-de Haas oscillations, as a function of the sheet carrier density. The dashed line indicates the bulk effective mass of ZnO. Adapted from Refs. 92 and 94.

**FIG. 22.** (Color online) (a) Three examples of ESR signal and the fitting using Eq. (4) to extract the resonance magnetic field and the transverse spin relaxation time. The full widths at half maximum from the fittings ($\Delta B$) are also schematically shown. (b) Transverse spin relaxation time as a function of resonance magnetic field with error bars obtained from the fitting results. The contribution of the Rashba effect estimated from Eq. (5) is also shown by the green curve. The arrows indicate the points shown in (a). Adapted from Ref. 95.

TABLE I: Physical parameters of ZnO. Those are lattice constants of $a$ and $c$ in Wurtzite structure, direct band gap ($E_g$), effective electron mass ($m^*$) normalized by bare electron mass ($m_0$), relative dielectric constant ($\varepsilon/\varepsilon_0$), exciton binding energy ($E_X$), and positions of conduction band ($E_C$) and valence band ($E_V$ ) edges.

| Parameter | $a$ (nm) | $c$ (nm) | $E_g$ (eV) | $m^*/m_0$ | $\varepsilon/\varepsilon_0$ | $E_X$ (meV) | $E_C$(eV) | $E_V$ (eV) |
|---|---|---|---|---|---|---|---|---|
| Value | 0.3250 | 0.5204 | 3.37 | 0.29 | 8.3 | 60 | 4.1 | 7.5 |

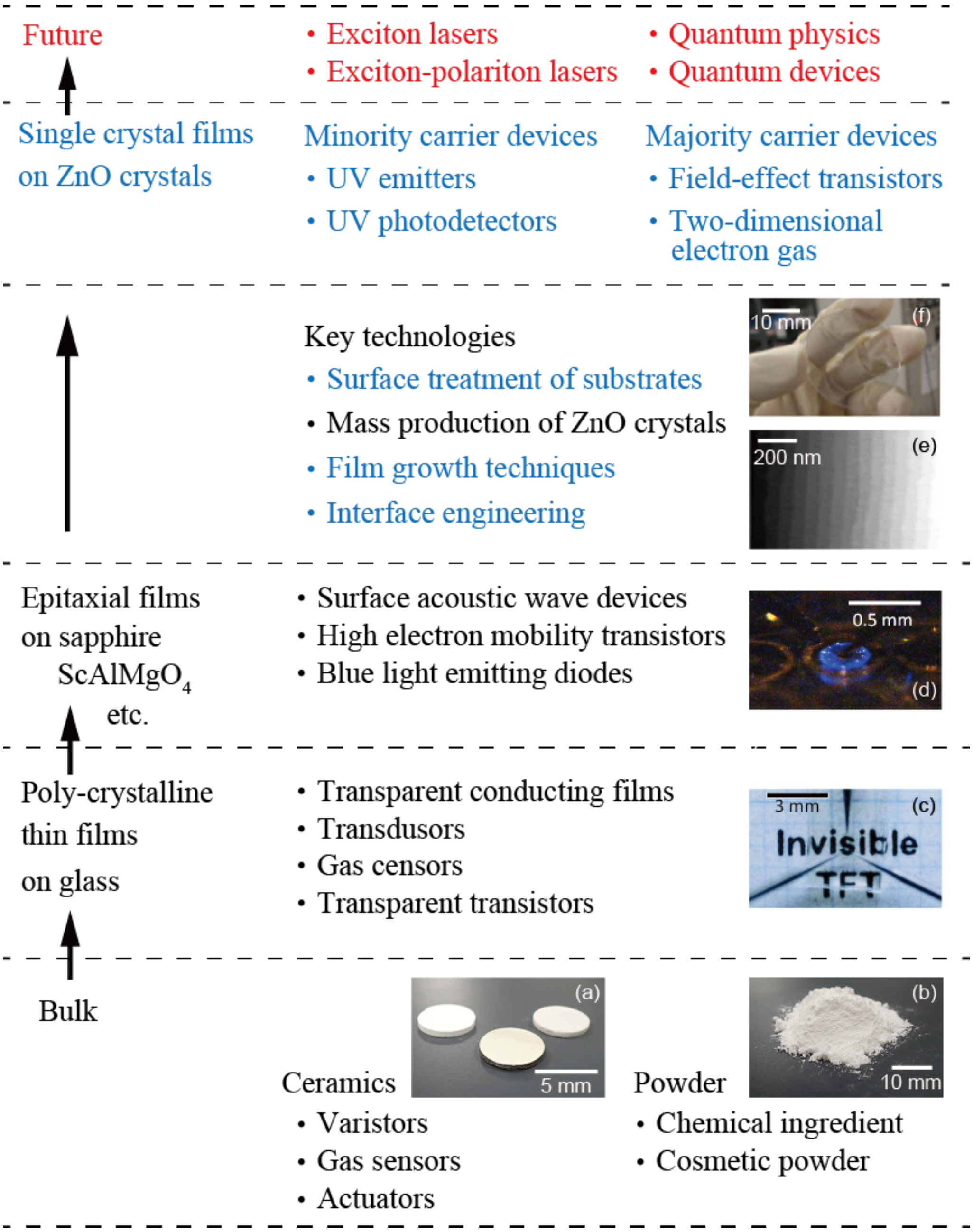

Fig. 1 Y. Kozuka *et al.*

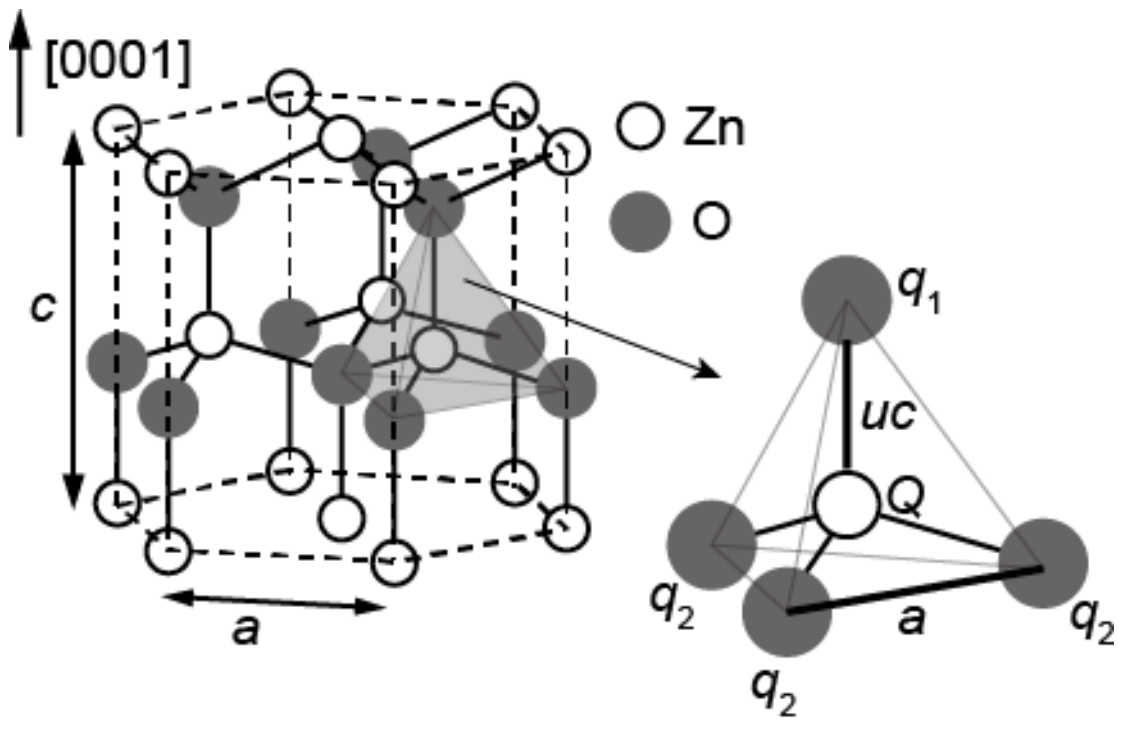

Fig. 2 Y. Kozuka *et al.*

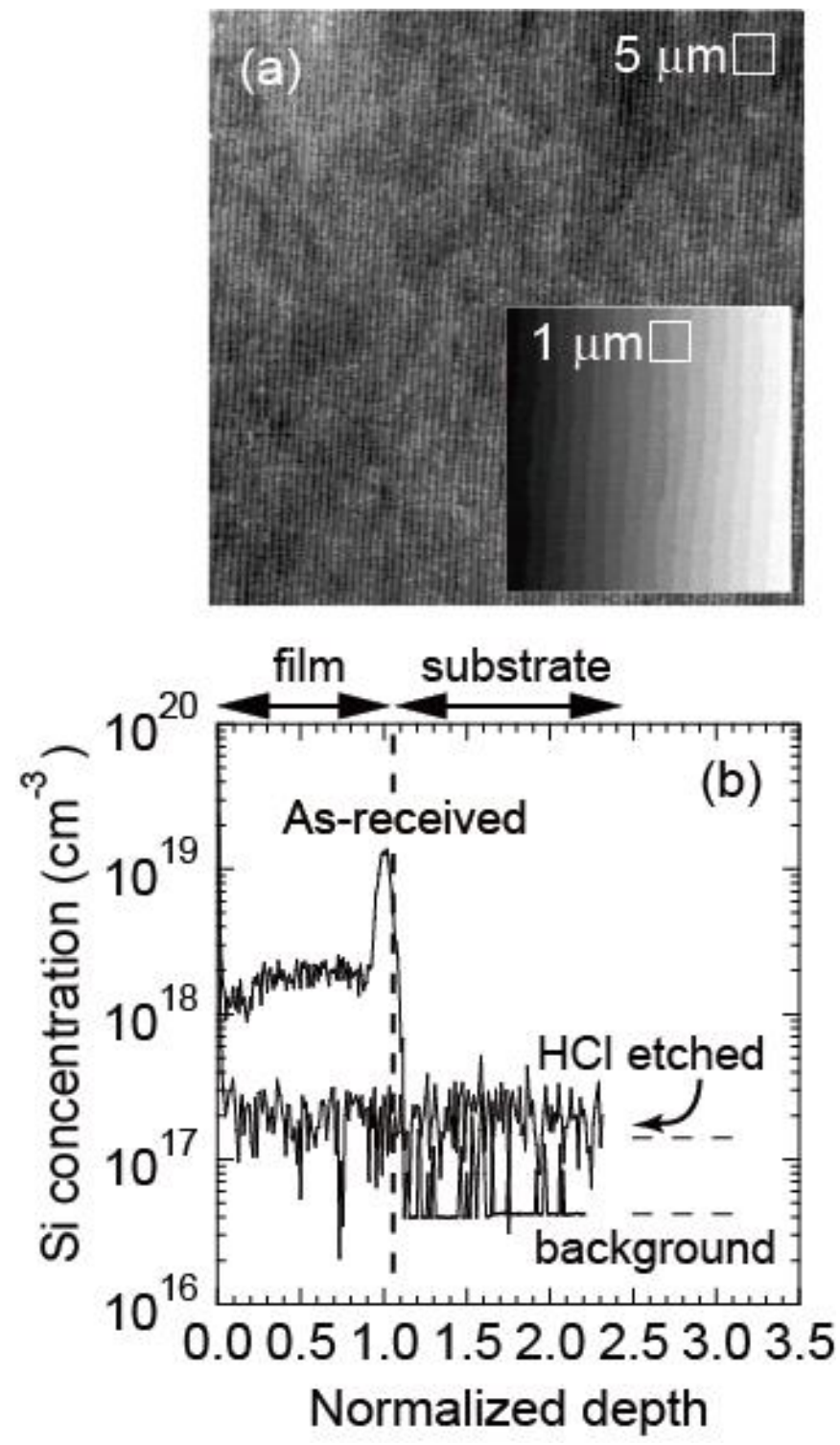

Fig. 3 Y. Kozuka *et al.*

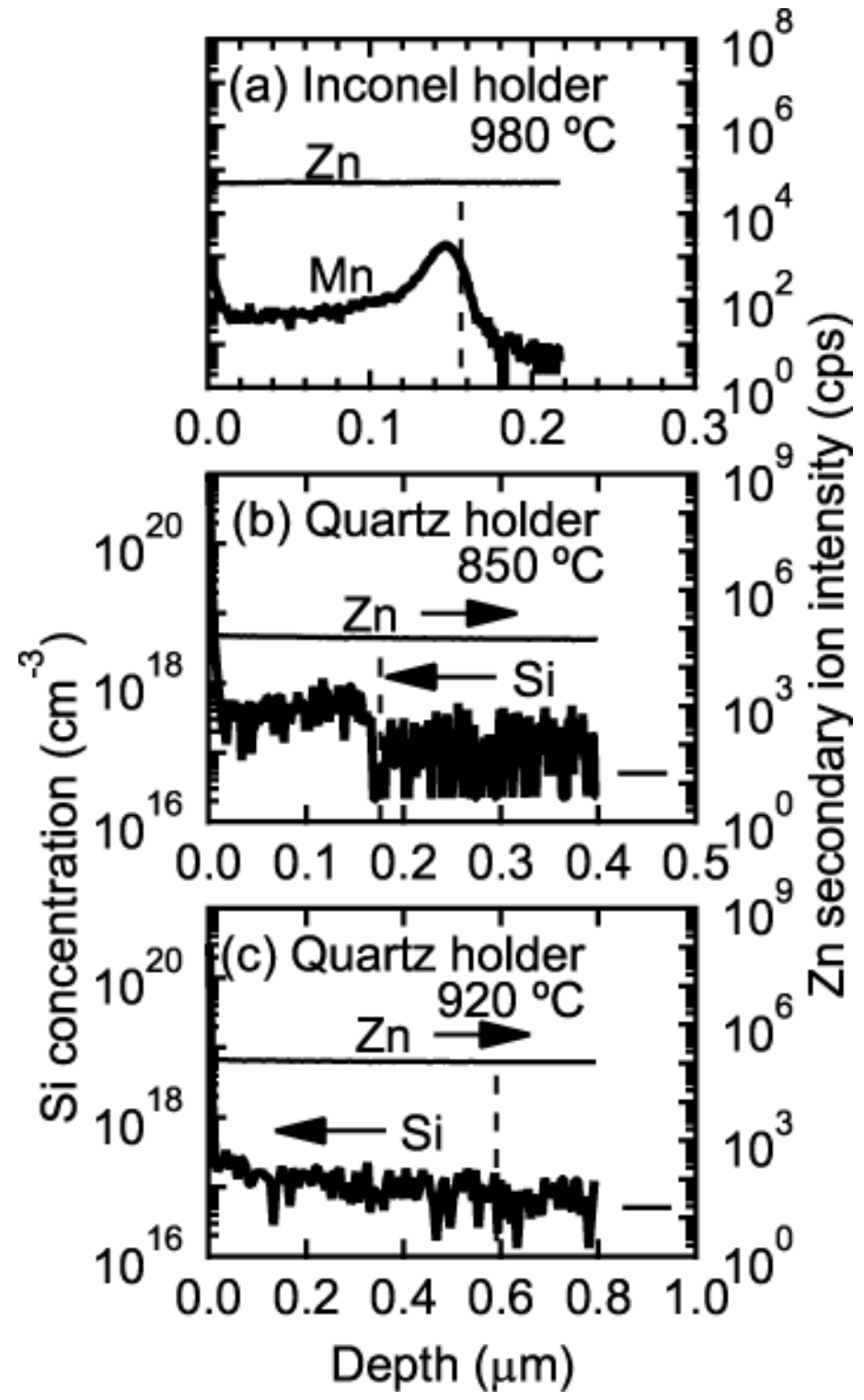

Fig. 4 Y. Kozuka *et al.*

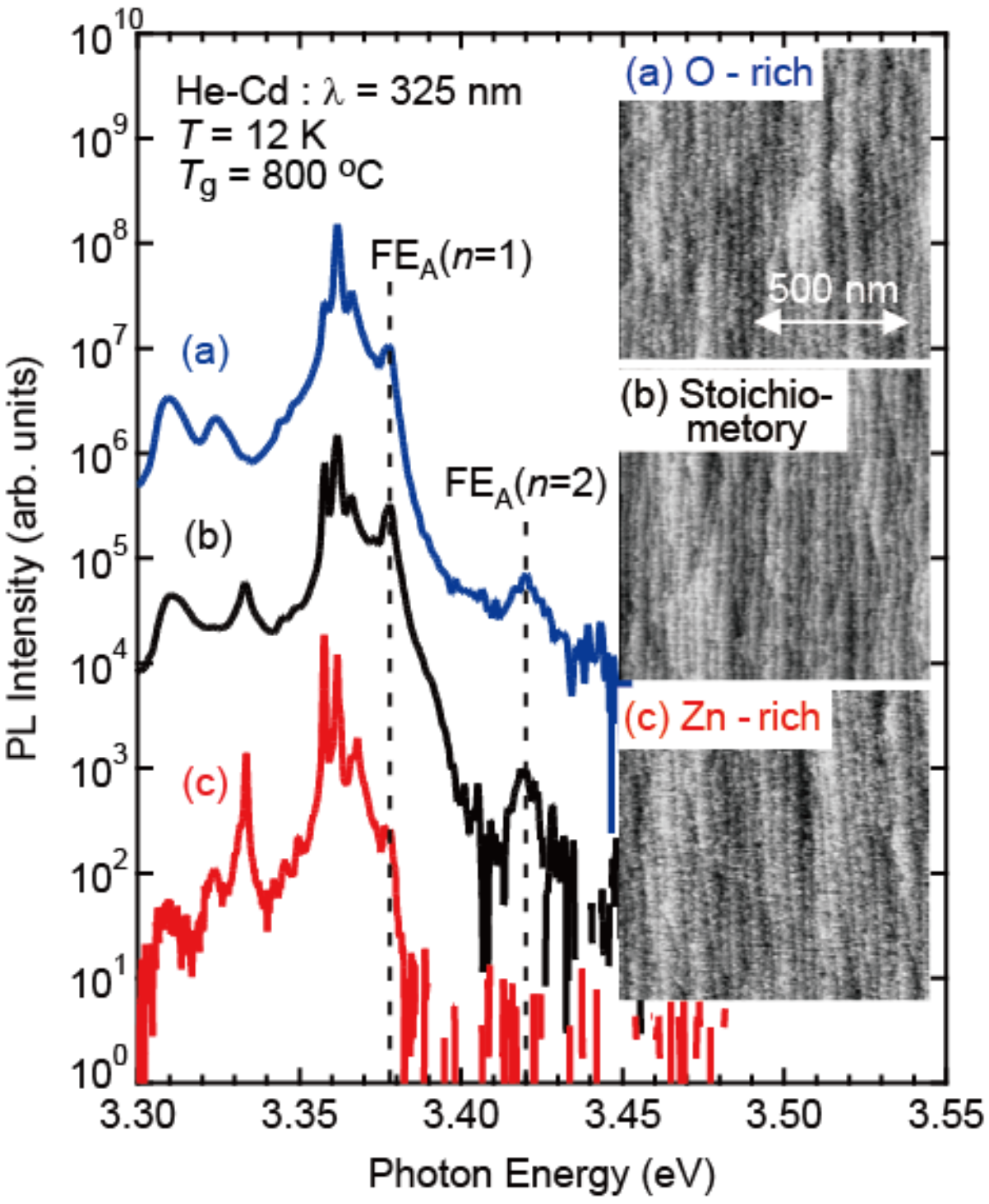

Fig. 5 Y. Kozuka *et al.*

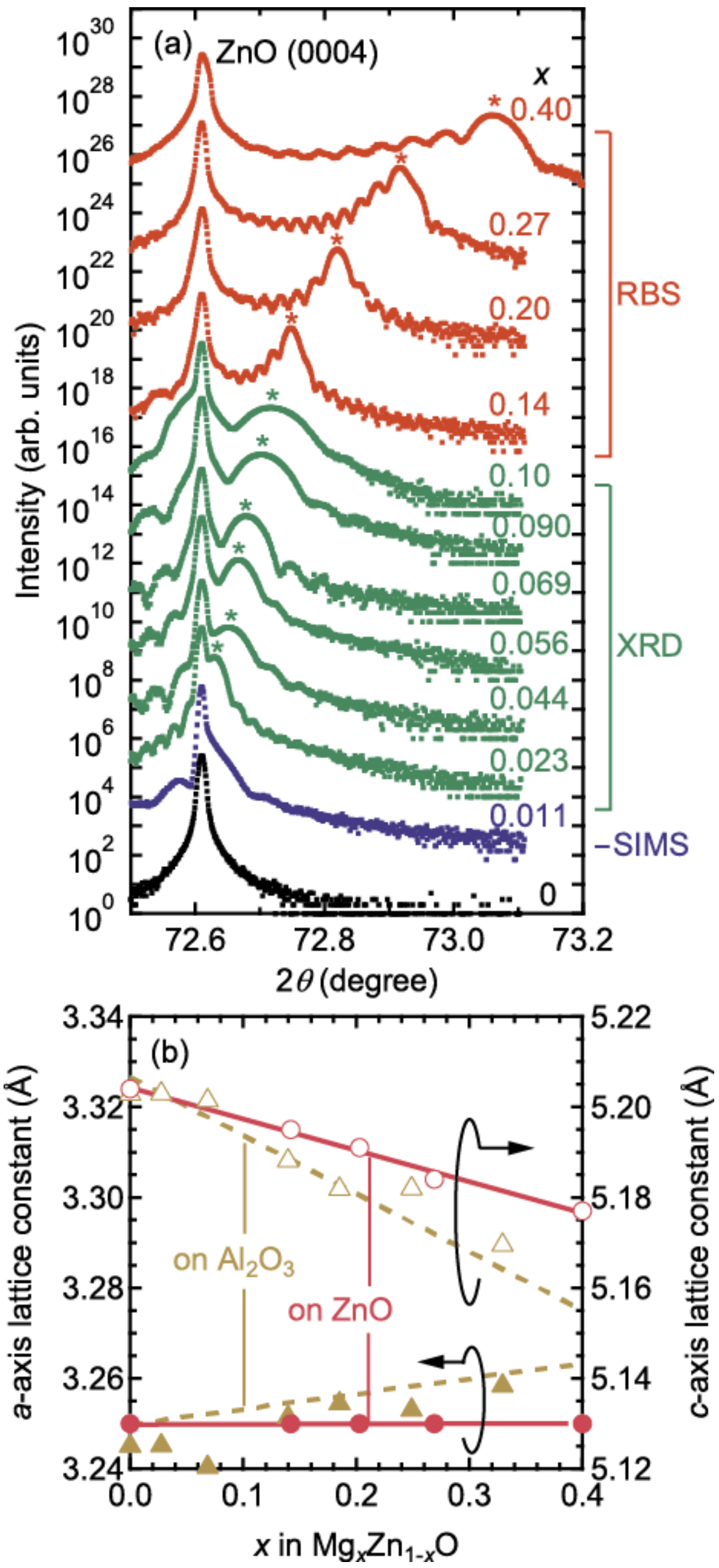

Fig. 6 Y. Kozuka *et al.*

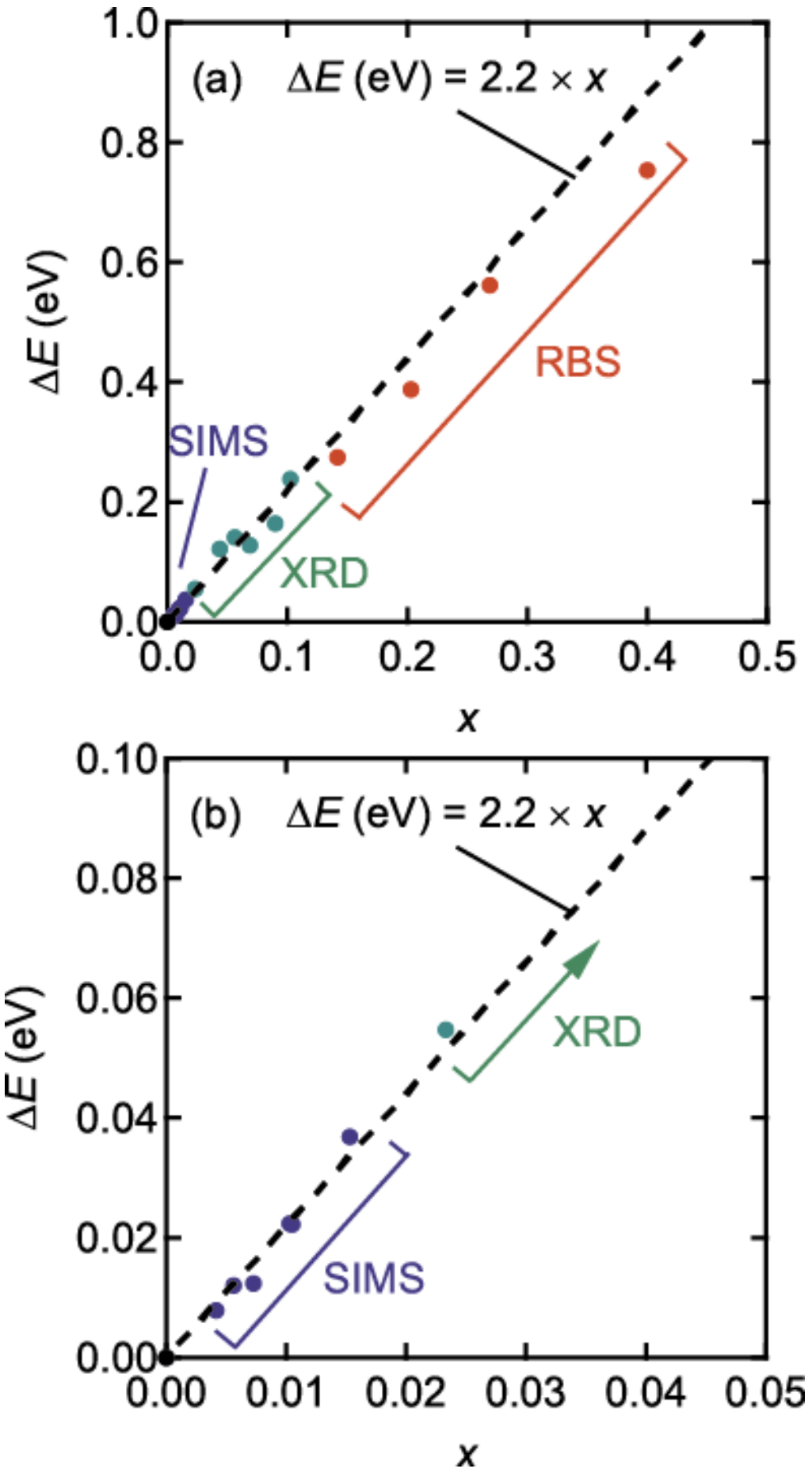

Fig. 7 Y. Kozuka *et al.*

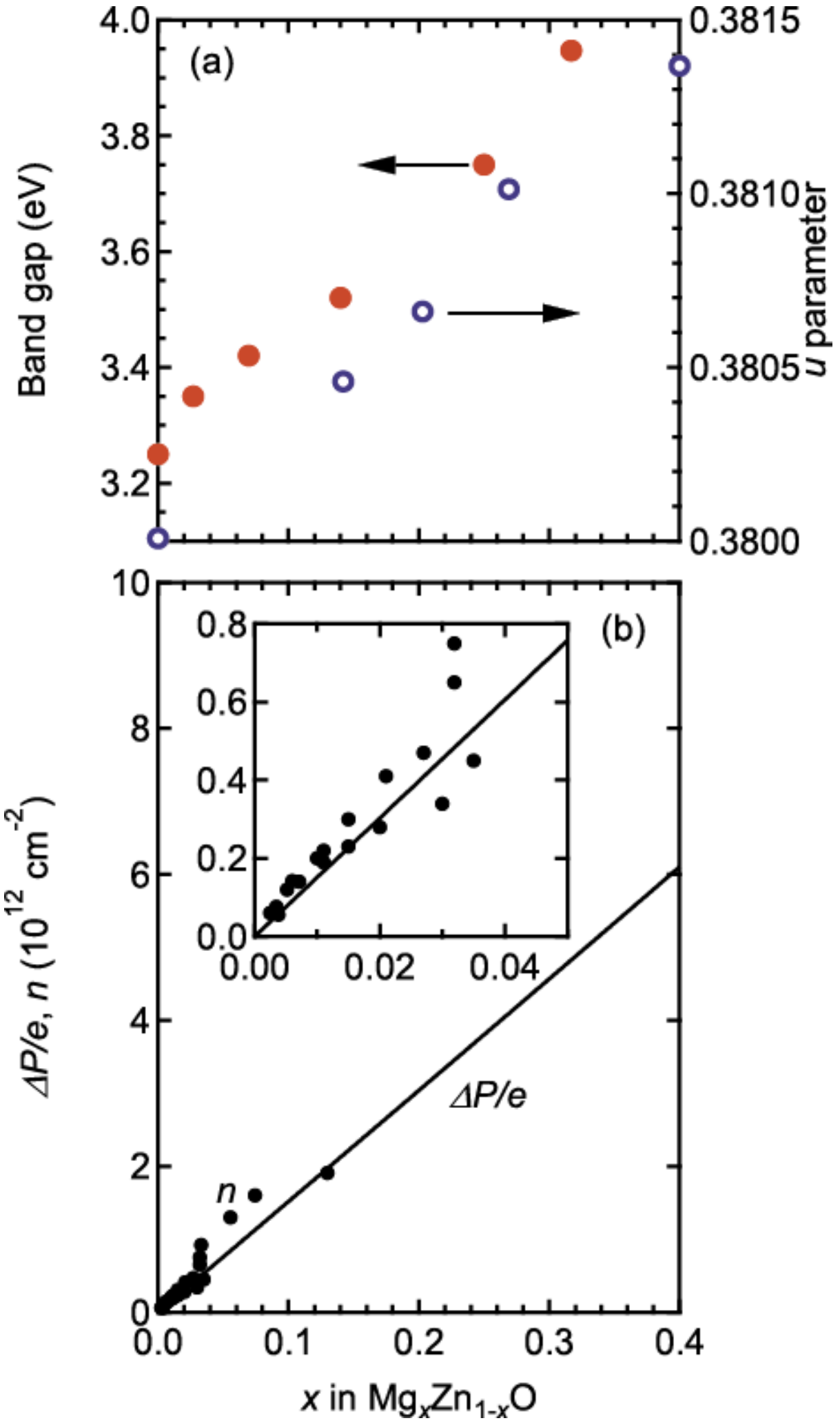

Fig. 8 Y. Kozuka *et al.*

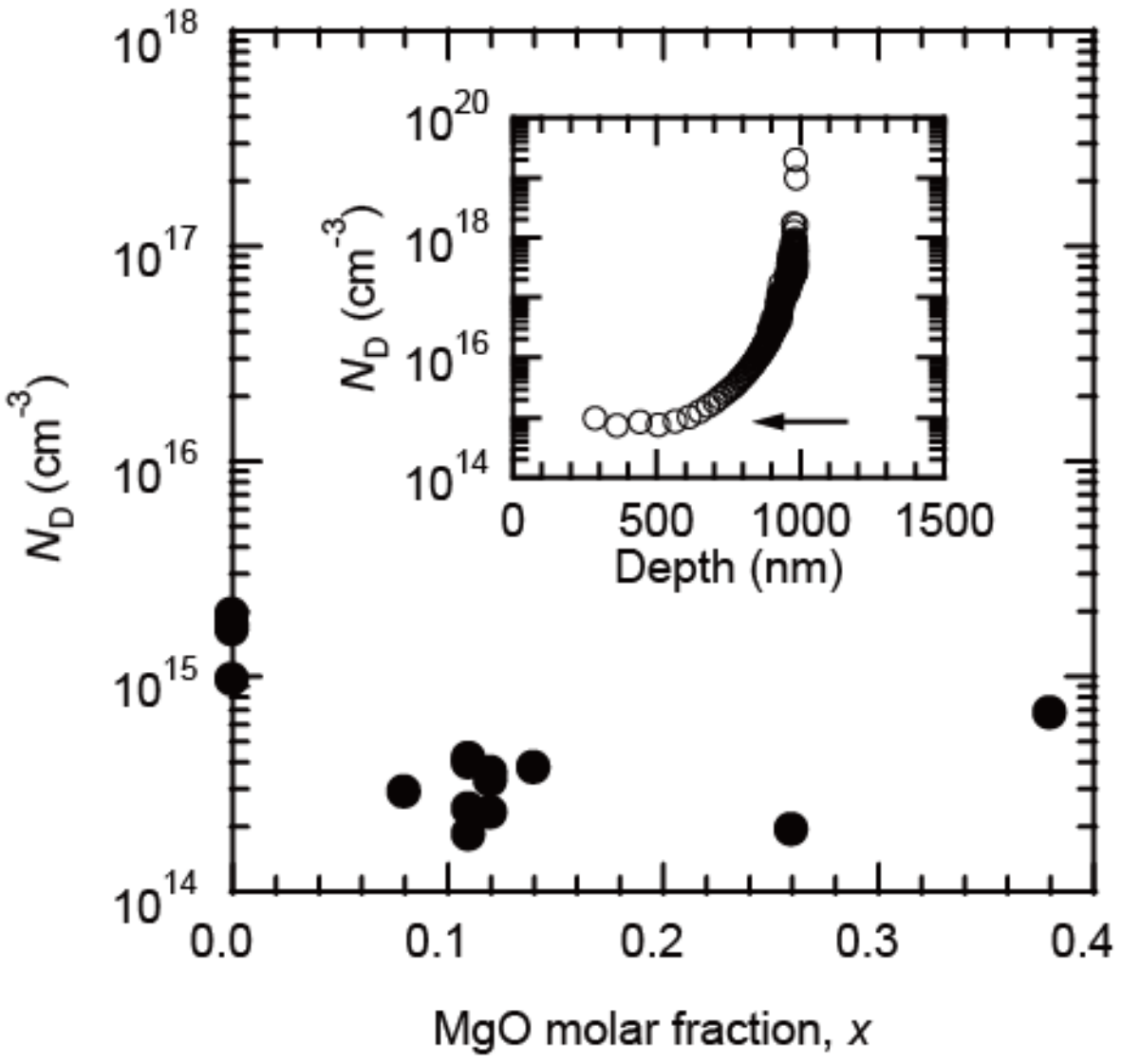

Fig. 9 Y. Kozuka *et al.*

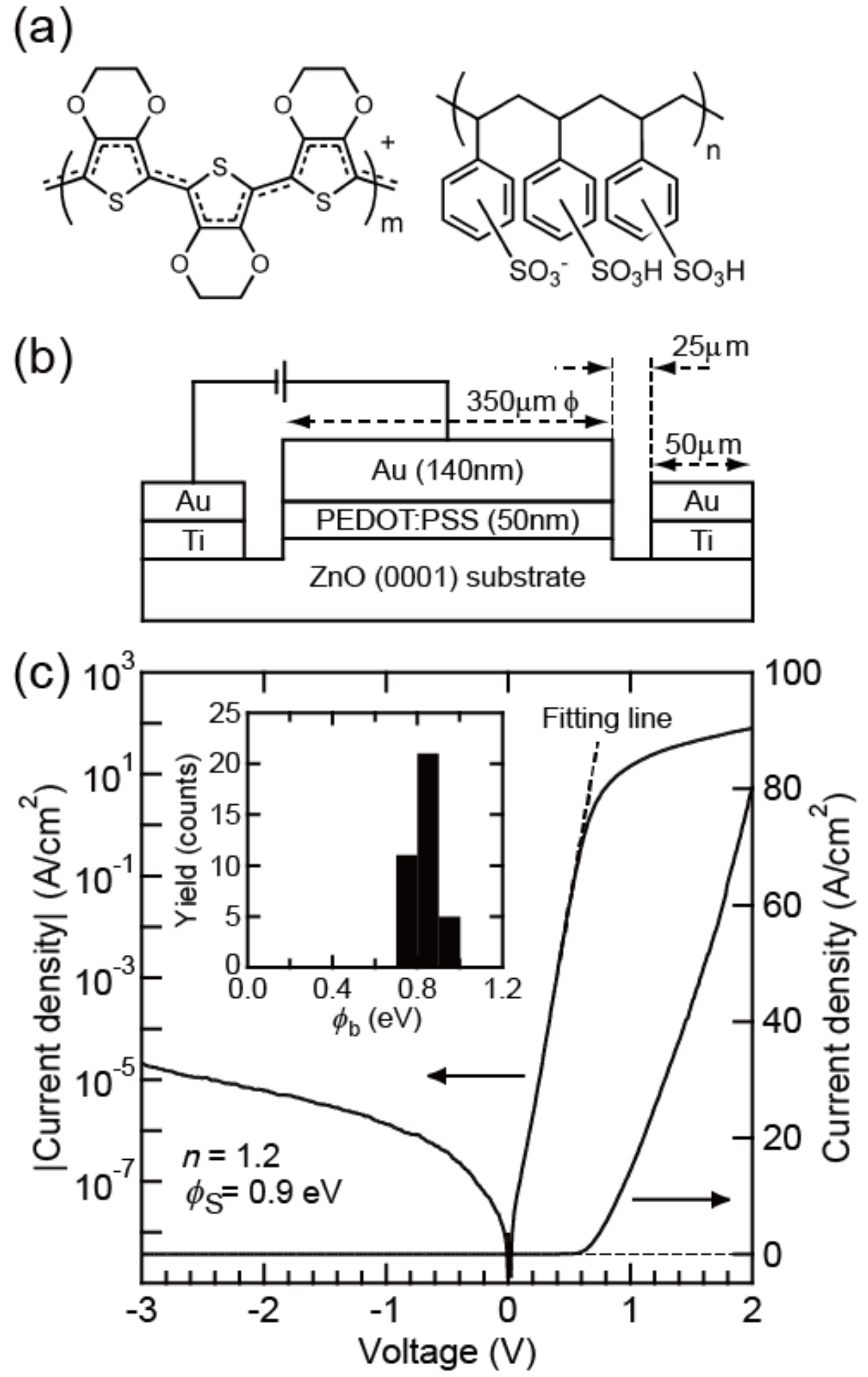

Fig. 10 Y. Kozuka *et al.*

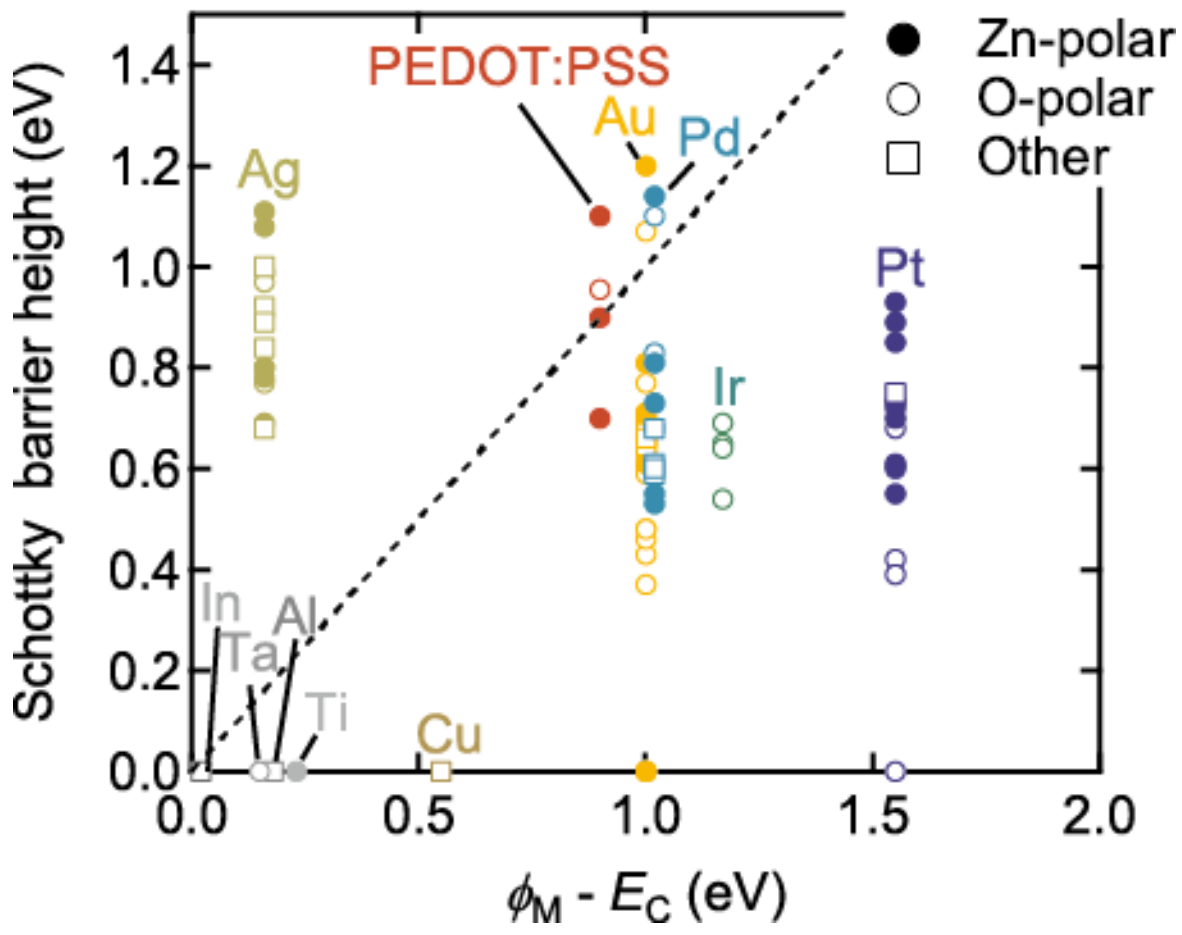

Fig. 11 Y. Kozuka *et al.*

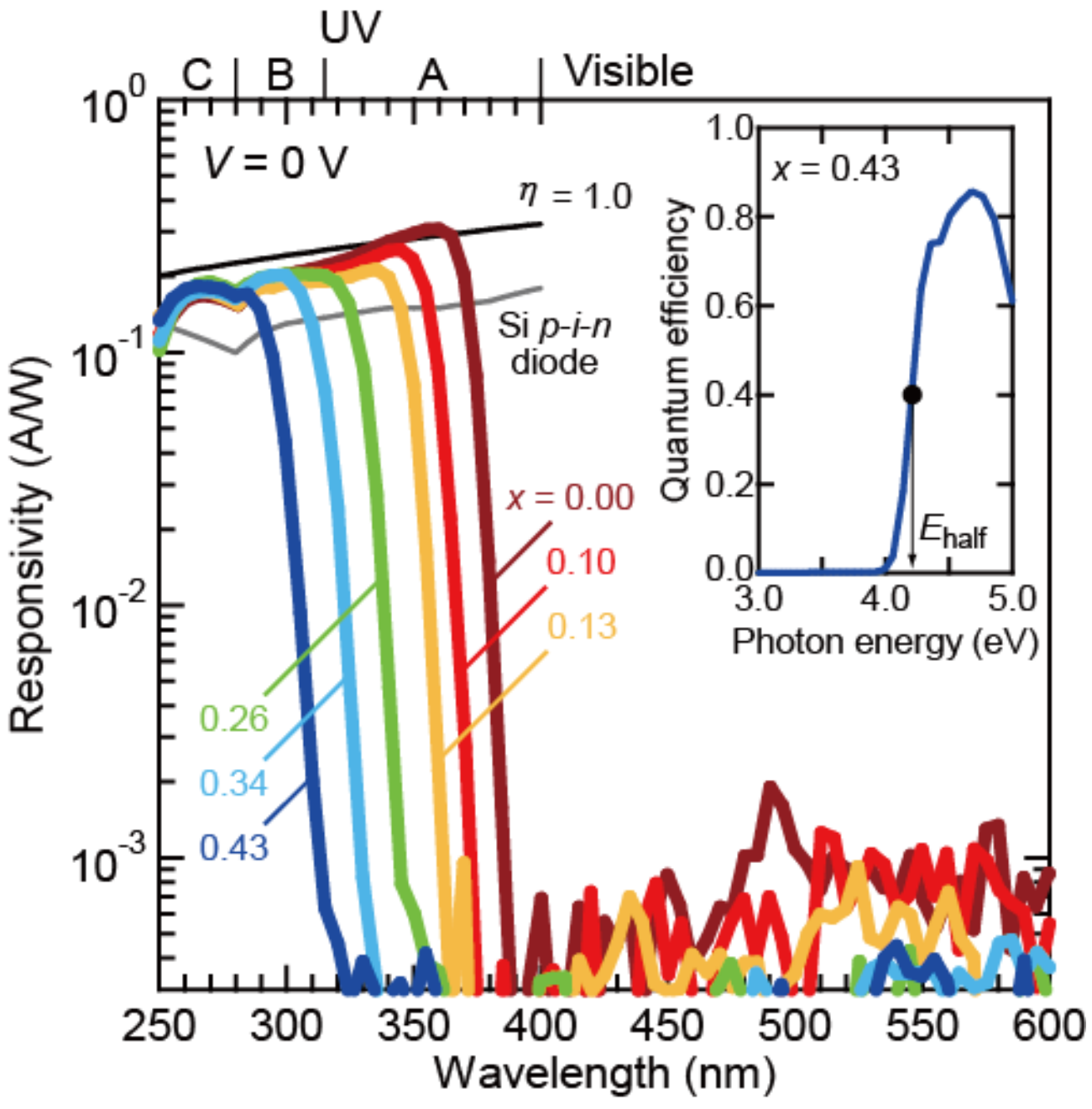

Fig. 12 Y. Kozuka *et al.*

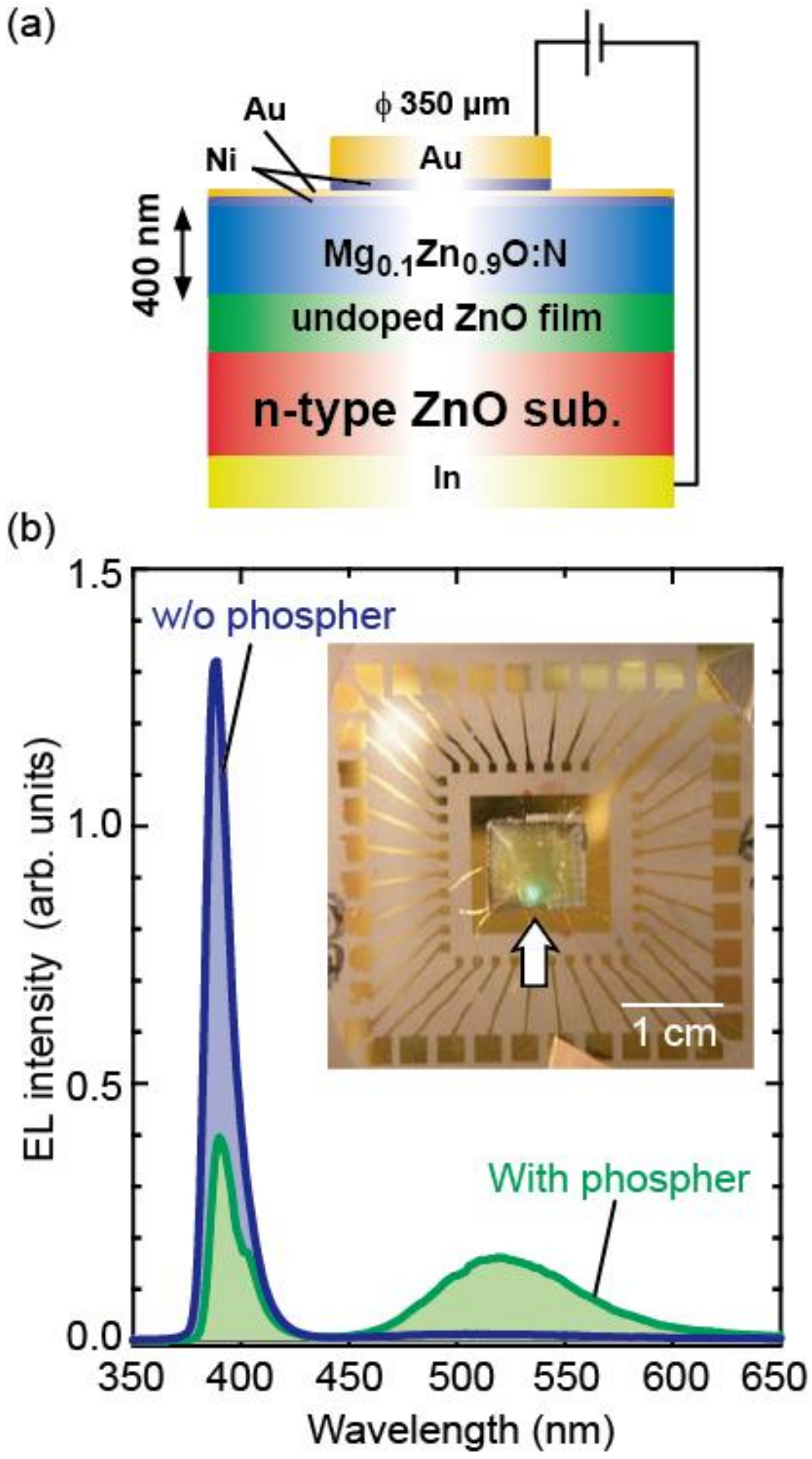

Fig. 13 Y. Kozuka *et al.*

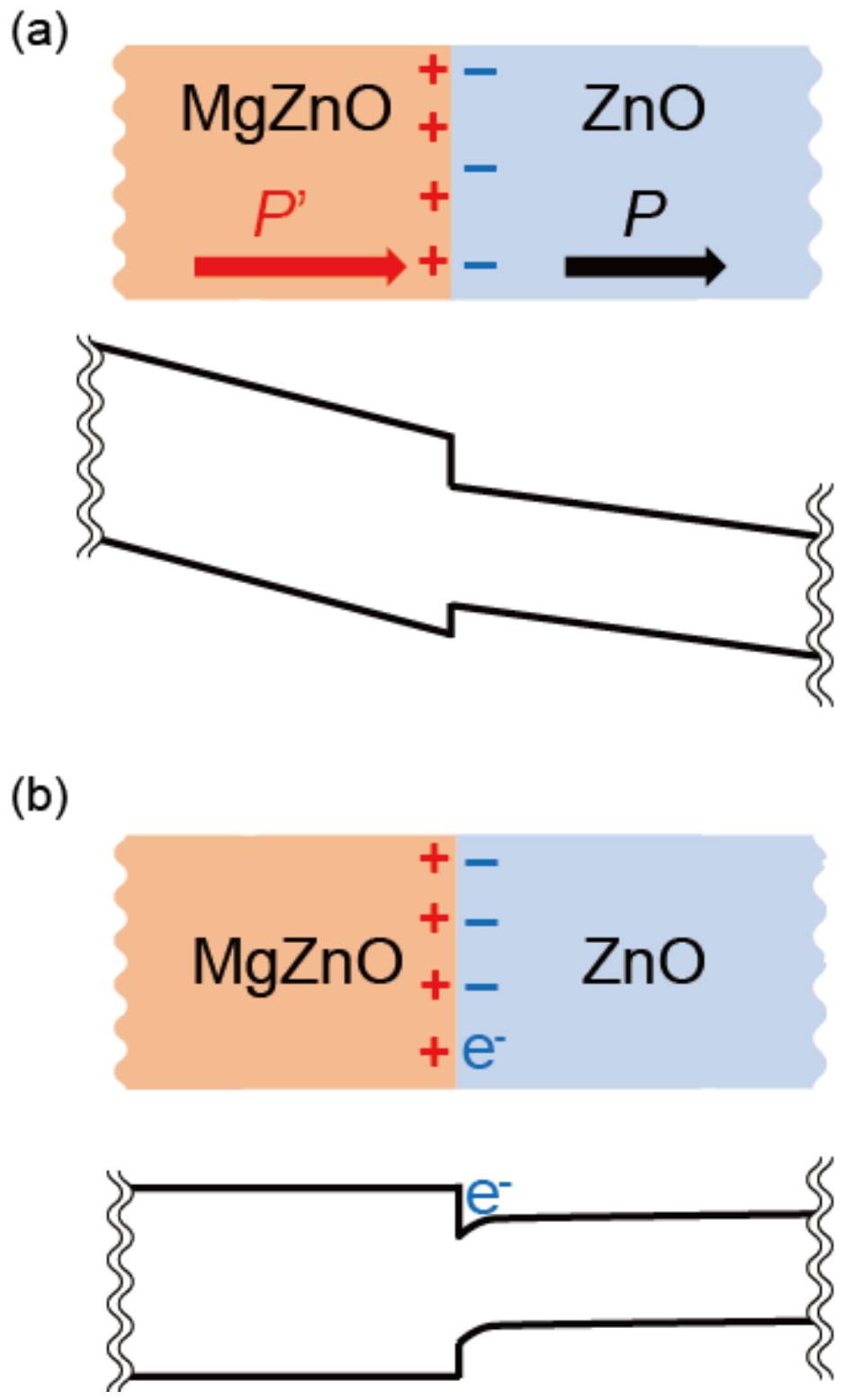

Fig. 14 Y. Kozuka *et al.*

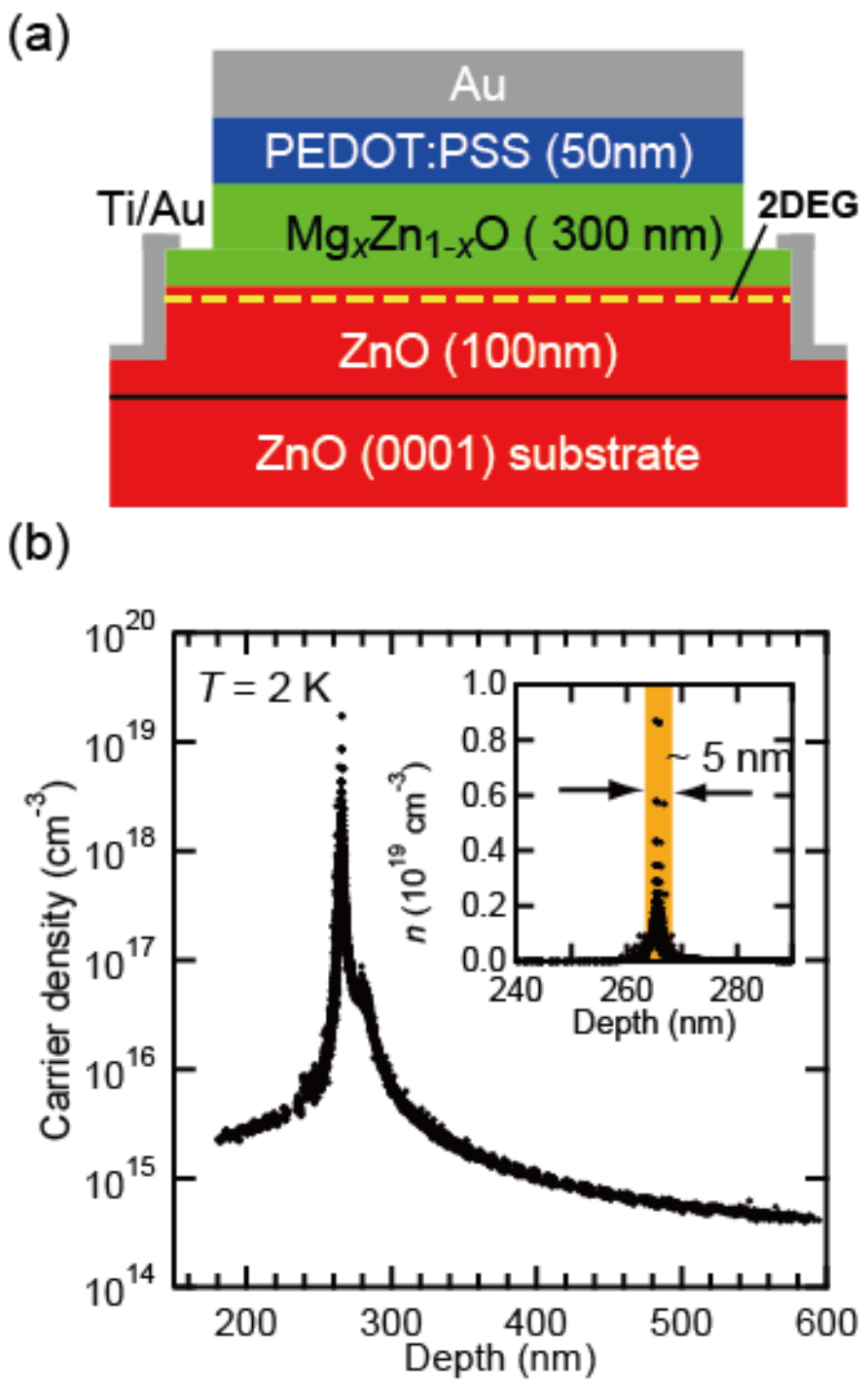

Fig. 15 Y. Kozuka *et al.*

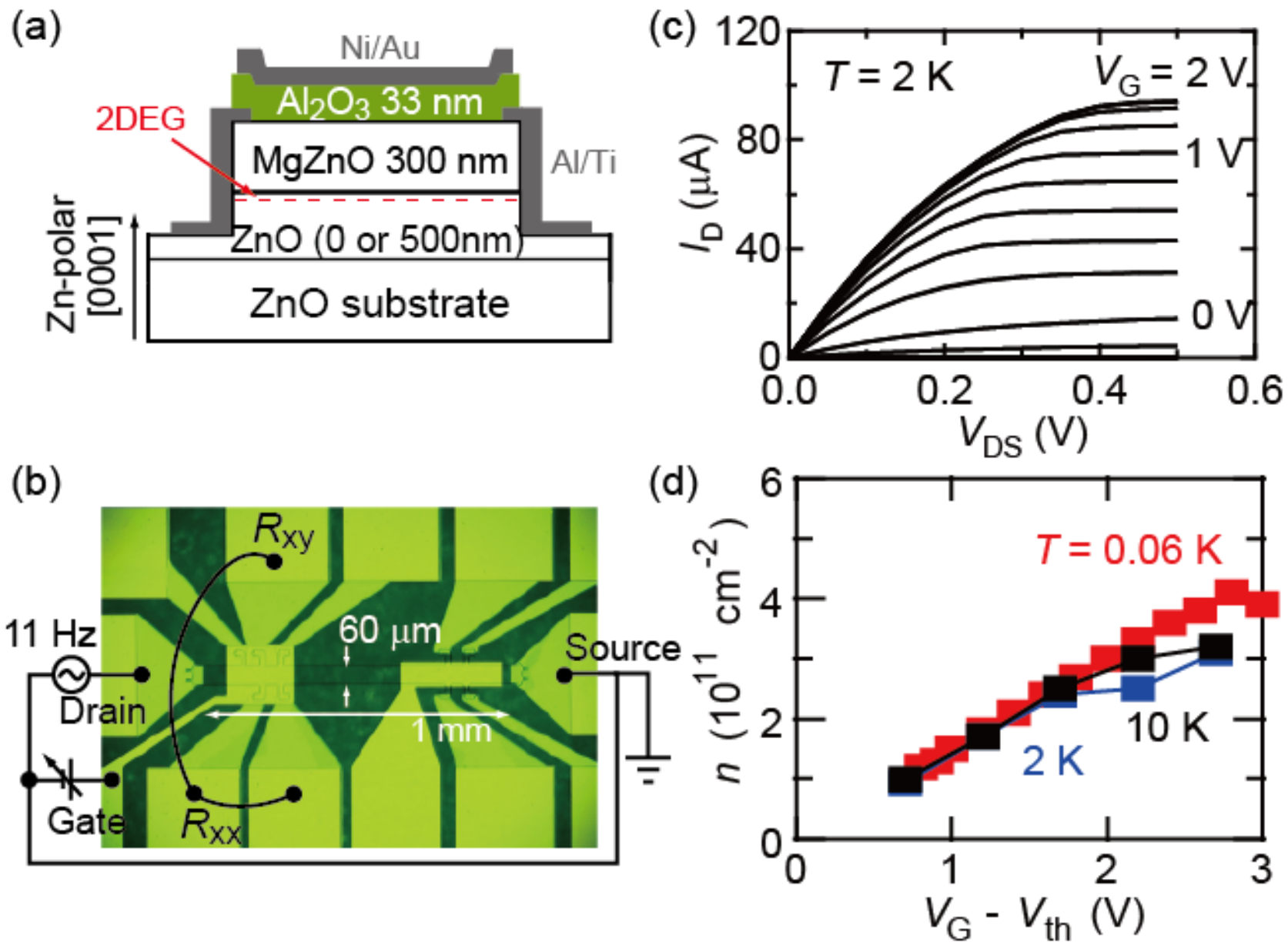

Fig. 16 Y. Kozuka *et al.*

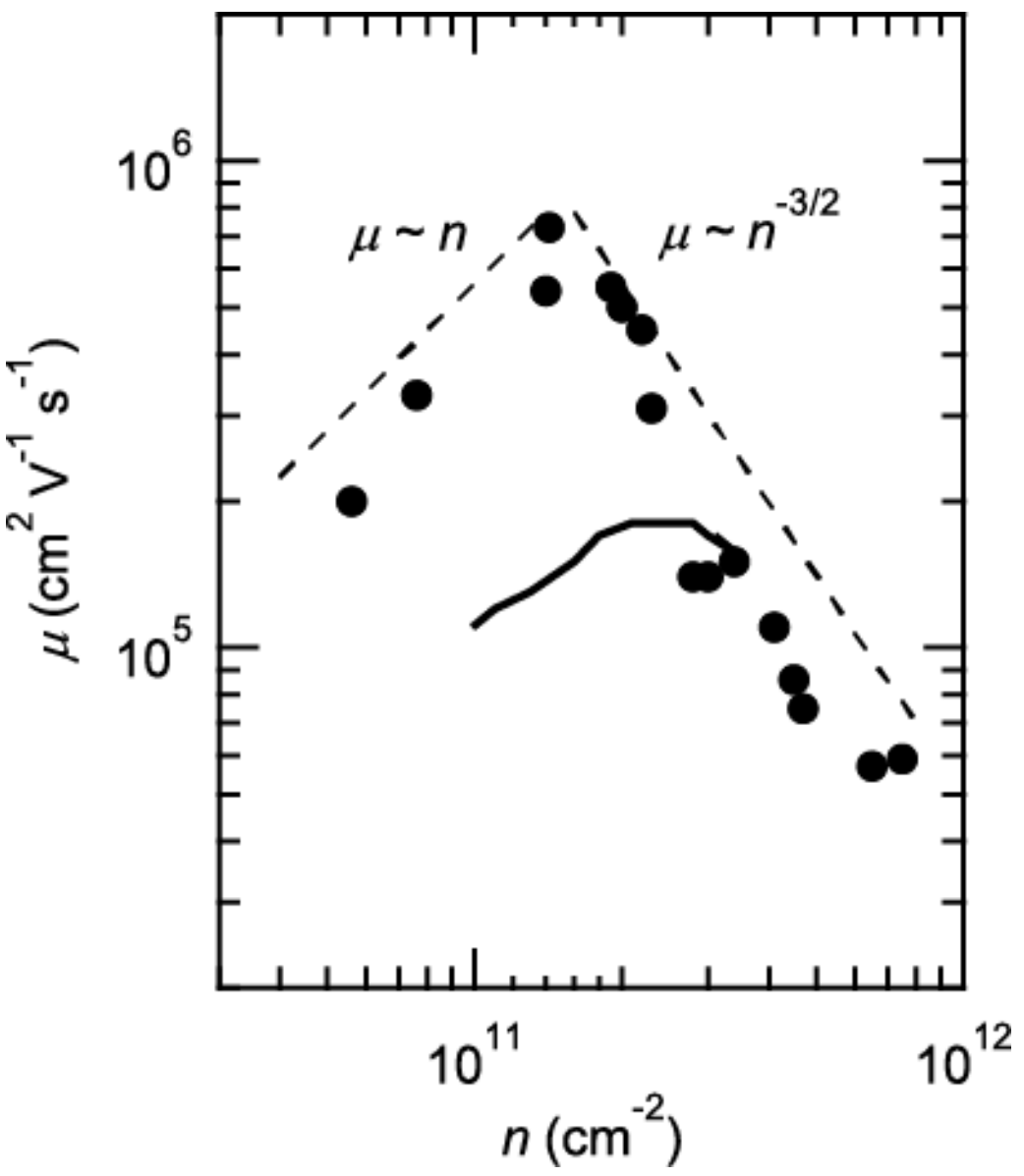

Fig. 17 Y. Kozuka *et al.*

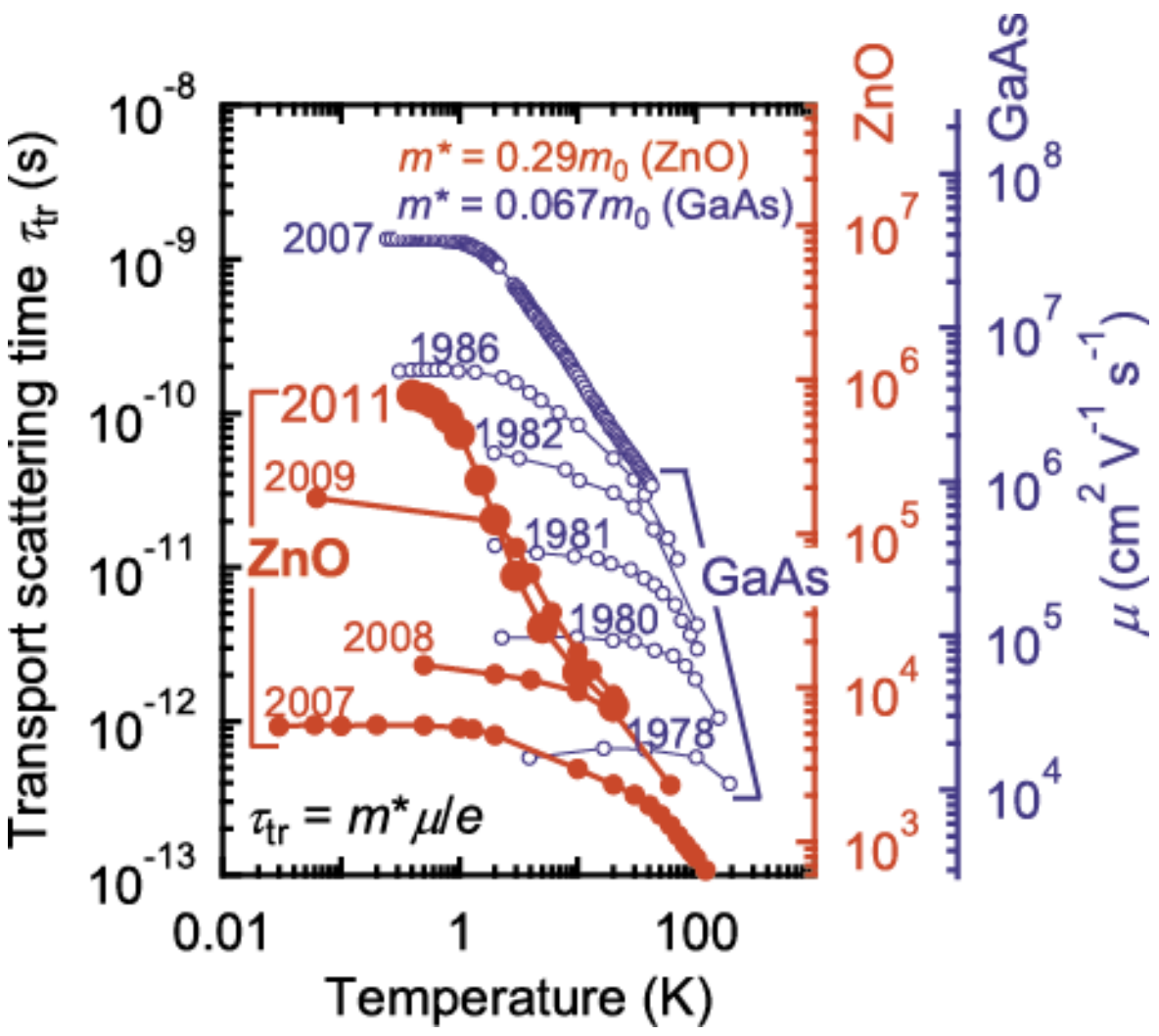

Fig. 18 Y. Kozuka *et al.*

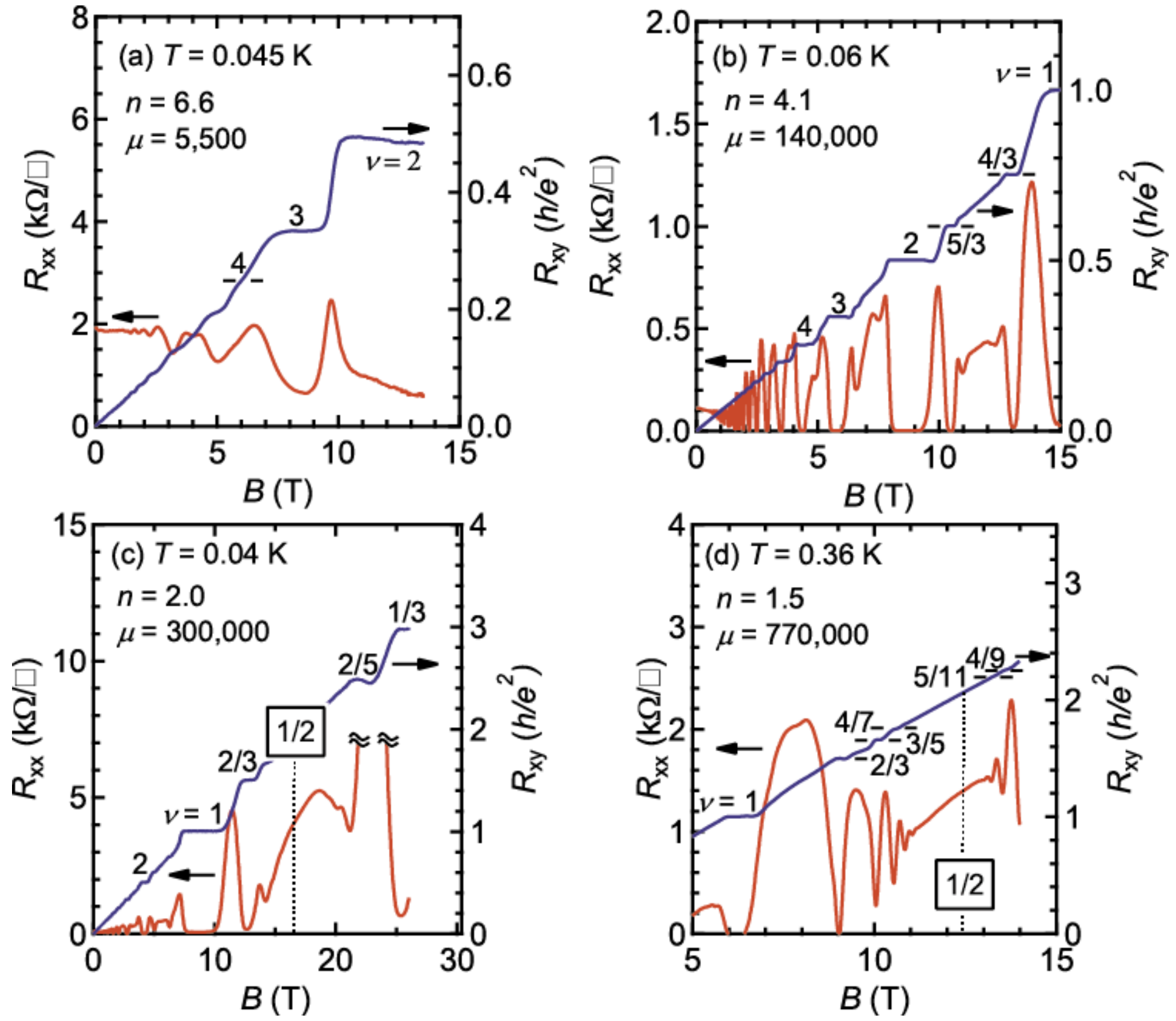

Fig. 19 Y. Kozuka *et al.*

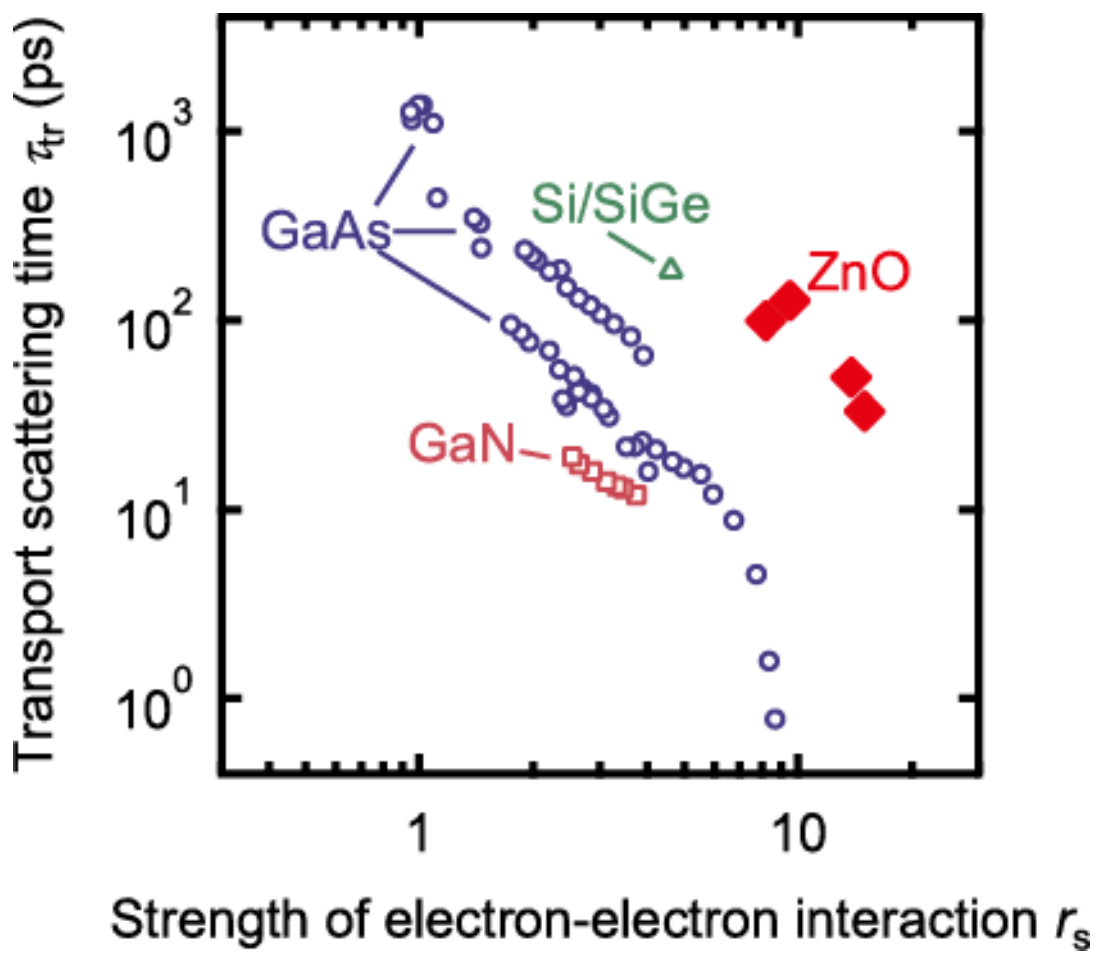

Fig. 20 Y. Kozuka *et al.*

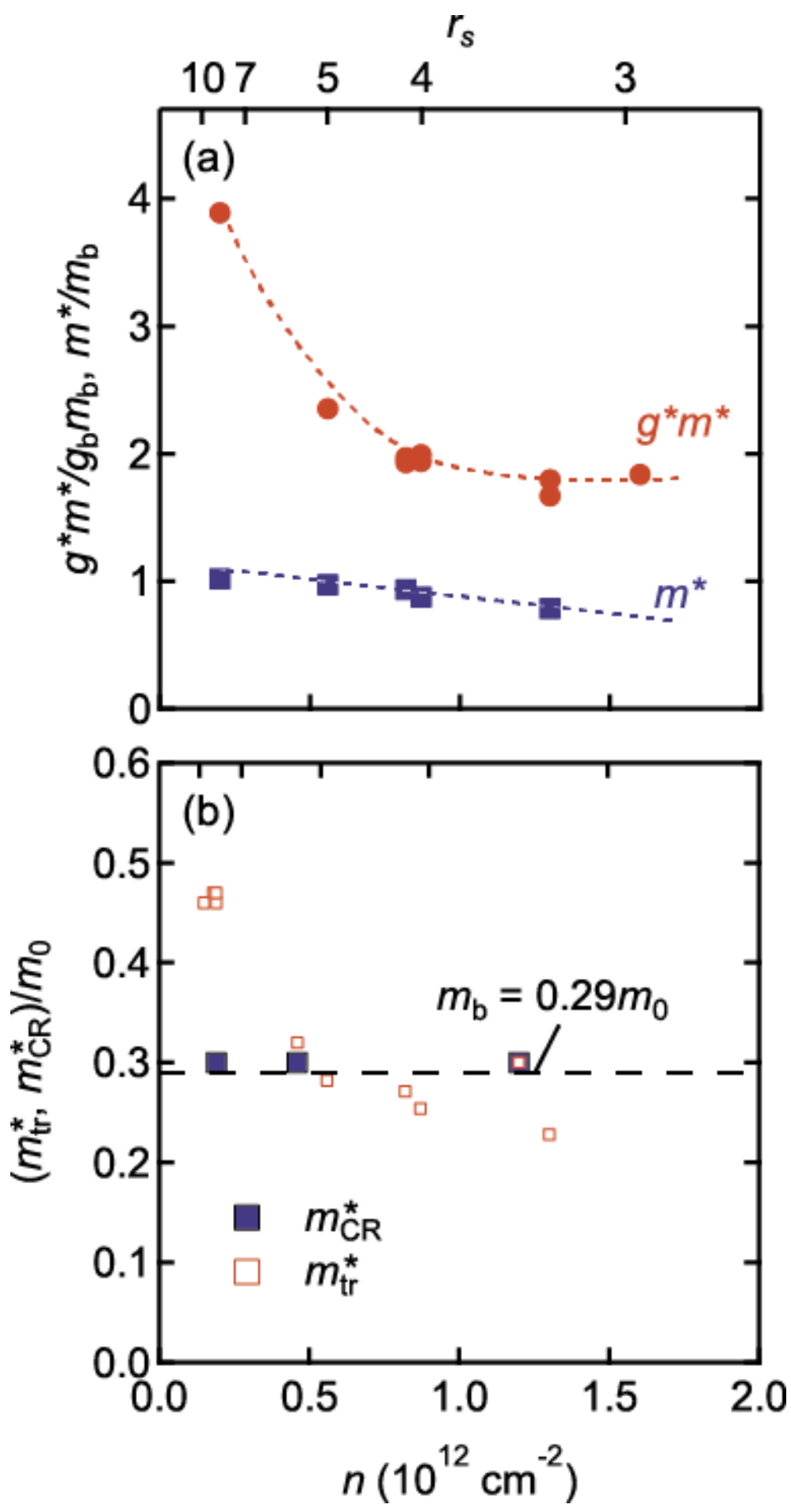

Fig. 21 Y. Kozuka *et al.*

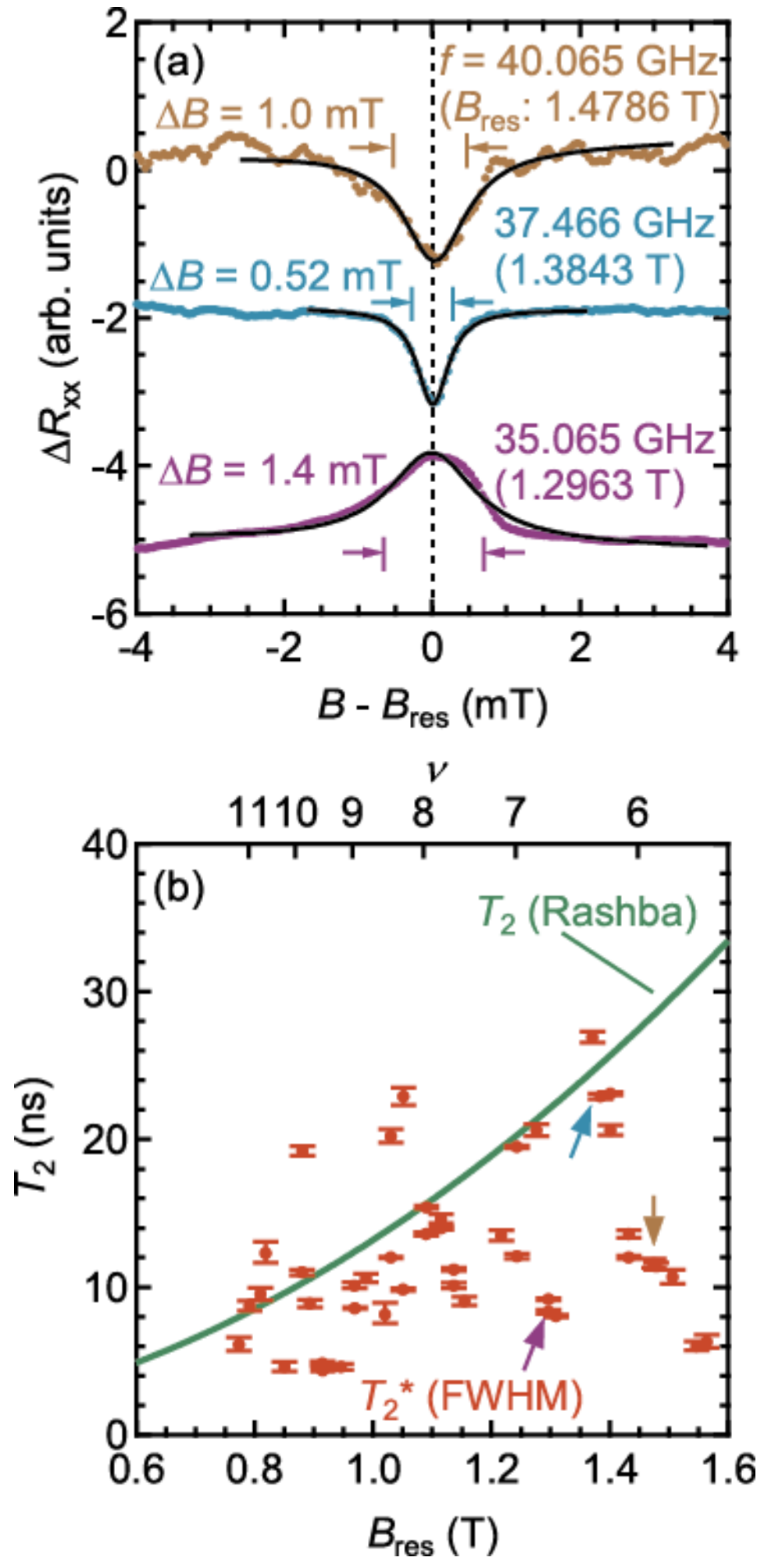

Fig. 22 Y. Kozuka *et al.*